\documentclass{article}
\usepackage[utf8]{inputenc}
\usepackage{graphicx}
\usepackage{amsmath,amssymb,amsthm}
\usepackage[ruled]{algorithm2e}
\usepackage{longtable}
\usepackage{authblk}
\usepackage[T1]{fontenc}
\usepackage[utf8]{inputenc}

\advance\textwidth48mm
\advance\hoffset-23mm
\advance\textheight32mm
\advance\voffset-15mm

\DeclareMathOperator*{\argmin}{arg\,min}

\newtheorem{example}{Example}
\newcommand{\cost}{\textsf{cost}}
\newcommand{\dist}{\textsf{dist}}

\newcommand{\swap}[2]{{#1}\overleftarrow{\to}{#2}}

\title{Quantum Circuit Transformation Based on Simulated Annealing and Heuristic Search}

\author[1,2]{Xiangzhen Zhou}
\author[2]{Sanjiang Li\thanks{sanjiang.li@uts.edu.au}}
\author[2]{Yuan Feng\thanks{yuan.feng@uts.edu.au}}

\affil[1]{State Key Lab of Millimeter Waves, Southeast University, Nanjing 211189, China}
\affil[2]{Centre for Quantum Software and Information,
Faculty of Engineering and Information Technology,
University of Technology Sydney, NSW 2007, Australia}

\date{June 2019}

\begin{document}
\maketitle
\begin{abstract}

Quantum algorithm design usually assumes access to a perfect quantum computer with ideal properties like full connectivity, noise-freedom and arbitrarily long coherence time. In Noisy Intermediate-Scale Quantum (NISQ) devices, however, the number of qubits is highly limited and quantum operation error and qubit coherence are not negligible. Besides, the connectivity of physical qubits in a quantum processing unit (QPU) is also strictly constrained. Thereby, additional operations like SWAP gates have to be inserted to satisfy this constraint while preserving the functionality of the original circuit. This process is known as quantum circuit transformation. Adding additional gates will increase both the size and depth of a quantum circuit and therefore cause further decay of the performance of a quantum circuit. Thus it is crucial to minimize the number of added gates. In this paper, we propose an efficient method to solve this problem. We first choose by using simulated annealing an initial mapping which fits well with the input circuit and then, with the help of a heuristic cost function, stepwise apply the best selected SWAP gates until all quantum gates in the circuit can be executed. Our algorithm runs in time polynomial in all parameters including the size and the qubit number of the input circuit, and the qubit number in the QPU. Its space complexity is quadratic to the number of edges in the QPU. Experimental results on extensive realistic circuits confirm that the proposed method is efficient and can reduce by 57\% on average the size of the output circuits when compared with the state-of-the-art algorithm on the most recent IBM quantum device viz. IBM Q20 (Tokyo).
\end{abstract}

\section{Introduction}

In Noisy Intermediate-Scale Quantum (NISQ) era, it is unrealistic to implement quantum error correction due to the strictly limited number of qubits \cite{NISQ}. 
This drawback brings huge challenge to quantum program compilation because the noise will have large impact on final circuits and may often make the results meaningless. Besides, the connectivity of qubits in an NISQ device is also limited. Only those neighbouring qubits can be coupled and only between them can two-qubit operations be implemented \cite{IBM_intro}. As a result, a large number of modifications must be done to adapt a quantum circuit to the real quantum devices. This process is termed as quantum circuit transformation \cite{6}, qubit mapping \cite{Q20}, qubit allocation \cite{7}, qubit routing \cite{initialmap} or qubit movement \cite{qubitmovement} in the literature. We call it quantum circuit transformation in this paper.

Quantum circuit transformation is an essential part for quantum circuit compilation. The main idea behind is to convert an ideal quantum circuit, in which full connectivity among qubits is assumed and noise is ignored, to a quantum circuit respecting constraints imposed by the NISQ devices \cite{6}. Usually this process will bring in a large number of auxiliary gates like SWAP gates and Hadamard gates which will in turn increase both the size and depth of the generated quantum circuit and sometimes make the error of the whole circuit unacceptable \cite{IBM_intro}. Hence, it is vital for the success of quantum computation to find an automated approach that can efficiently transform any input quantum circuit into one that respects the physical constraints imposed by the NISQ devices with a small overhead in terms of the size, depth or error.

The quantum circuit transformation problem can be reduced to token swapping or template matching in graph theory \cite{token,matching}. Unfortunately, both of these problems are NP-hard \cite{6}.
Hence, designing algorithms to solve the quantum circuit transformation problem while making trade off between time consuming and the quality of results has brought lots of interest in both the quantum computing community and the  integrated circuits community \cite{IBM_intro}.

There are currently three major approaches to the quantum circuit transformation problem. The first one is to use heuristic search to construct the output quantum circuit step by step from the original input quantum circuit \cite{Q20,initialmap,Astar,7,13}. Usually, these search algorithms need an initial mapping as the input, and it can be set arbitrarily or via some greedy methods \cite{Astar,initialmap,11}. Recently, a novel reverse traversal technique is proposed in \cite{Q20} to choose the initial mapping with the consideration of the whole circuit. The second approach is to utilize unitary matrix decomposition algorithms to reconstruct a quantum circuit from scratch while preserving the functionality of the input circuit \cite{Ud1,Ud2}. The third one is to convert the quantum circuit transformation problem to some existing problems like AI planning and constraint programming and use ready-made tools for these problems to find acceptable results \cite{planning,planning2}.

In this paper, we follows the first approach. Our main contributions are listed as follows. First, we propose a simulated annealing based algorithm to find a near-optimal initial mapping for the input circuit. Second, we design a flexible heuristic cost function to evaluate the possible operations that may be applied to transform the current circuit. The heuristic function supports weight parameters to reflect the variable influence of gates in different layers. Third, a heuristic search algorithm  with a novel selection mechanism is designed, where in each step of the search process, instead of selecting the operation with minimum cost to apply, we look one step ahead and select the operation which has the best consecutive operation to apply. In this way, the algorithm is able to avoid the local minimum effectively. Fourth, a pruning mechanism is introduced to reduce the size of search space and ensure the program terminates in reasonable time.

Note that the look-ahead mechanism has already been introduced in the heuristic cost function during the search process in existing works like \cite{Astar,Q20}. However, we adopt in this paper a double look-ahead mechanism: in addition to looking ahead at subsequent layers when defining the cost function, we also look ahead (at grandchild states) in finding the state with minimal cost in order to make the best transformation. Thanks to this novel idea, the proposed algorithm is able to find a better solution with less circuit size within acceptable running time. Experimental results on extensive realistic circuits show that our algorithm is efficient and, when compared with the state-of-the-art algorithms \cite{Astar,Q20}, can reduce on average the size of the output circuits by above 10\% on IBM QX5 and above 55\% on IBM Q20.

The remainder of this paper is organized as follows: some background knowledge about quantum computation is given in Section~\ref{sec:background}, and the quantum circuit transformation problem is formally defined in Section~\ref{sec:problem}. Section~\ref{sec:algorithm} is devoted to the detailed description of our proposed algorithm. We report experimental results in Section~\ref{sec:benchmark} and conclude the paper in Section~\ref{sec:conclusion}.

\section{Background}
\label{sec:background}

\begin{figure}[t]
	\centering
	\includegraphics[width=0.4\textwidth]{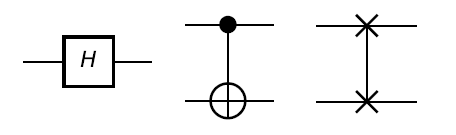}
	\caption{Hadamard, CNOT and SWAP gate (from left to right).}
	\label{fig:gate symbol}
\end{figure}

In classical computation, information is stored in memory in the form of binary digits, i.e., bits. The quantum counterpart of bit, called qubit, has two basis states denoted by $\left| 0 \right\rangle$ and $\left| 1 \right\rangle$, respectively. Different from a classical bit, a qubit $\left| \psi \right\rangle$ can be in a linear combination of basis states \cite{1}, i.e.,
\begin{equation}\label{eq:superposition}
    \left| \psi \right\rangle = \alpha\left| 0 \right\rangle + \beta\left| 1 \right\rangle
\end{equation}
where ${\left| \alpha  \right|^2} + {\left| \beta  \right|^2} = 1$.
Information processing or computation is realized by applying quantum gates on qubits. Typical gates which we are concerned with in this paper are Hadamard gate H, CNOT gate and SWAP gate, depicted in Fig.~\ref{fig:gate symbol}. H is a single-qubit gate which can evenly mix the basis states to produce a superposed one. CNOT and SWAP are both two-qubit gates, i.e., they operate on two qubits. A CNOT gate flips the target qubit (indicated graphically with $\oplus$) if and only if the control qubit (indicated graphically with a black dot $\bullet$) is in state $\left| 1 \right\rangle$, while a SWAP gate exchanges the states of the two qubits operated.

Quantum circuits are the most commonly used model to describe quantum algorithms, which consist of input qubits, quantum gates, measurements and classical registers \cite{2}. 
However, as far as quantum circuit transformation is concerned, only input qubits and quantum gates are relevant. Thus in this paper, a quantum circuit is simply represented as a pair $\left( {Q,C} \right)$, where $Q$ is the set of involved qubits and $C$ a sequence of quantum gates. 
For a generic quantum circuit to be executed in a real quantum processing unit (QPU), two more steps have to be taken: 
\begin{itemize}
	\item \emph{Compilation} process. As only limited quantum operations are available in a QPU, quantum gates in the circuit must be decomposed into elementary gates first \cite{10,compile}. In this paper, we take single-qubit and CNOT gates as elementary gates as they are universal to implement any quantum circuit and supported by, say, IBM QX architectures.
	
	\item \emph{Transformation} process. Qubits in a real QPU are typically laid out in a fixed topology and CNOT gates can only be applied on neighbouring qubits. Such a connectivity topology can be described by an \emph{architecture graph} or \emph{coupling graph} \cite{6} which is a 
	directed graph with each node representing a qubit in the QPU. A quantum circuit consisting of only single-qubit and CNOT gates is said to \emph{respect} the QPU constraint if for every 	
	CNOT gate in the circuit, there is a directed edge in the architecture graph from the control qubit to the target qubit. The transformation process is then to convert a quantum circuit (say, those obtained from the above compilation process) into one that respects the QPU constraint so that it can be executed on the QPU.
\end{itemize}

\begin{figure}[t]
	\centering
	\includegraphics[width=0.5\textwidth]{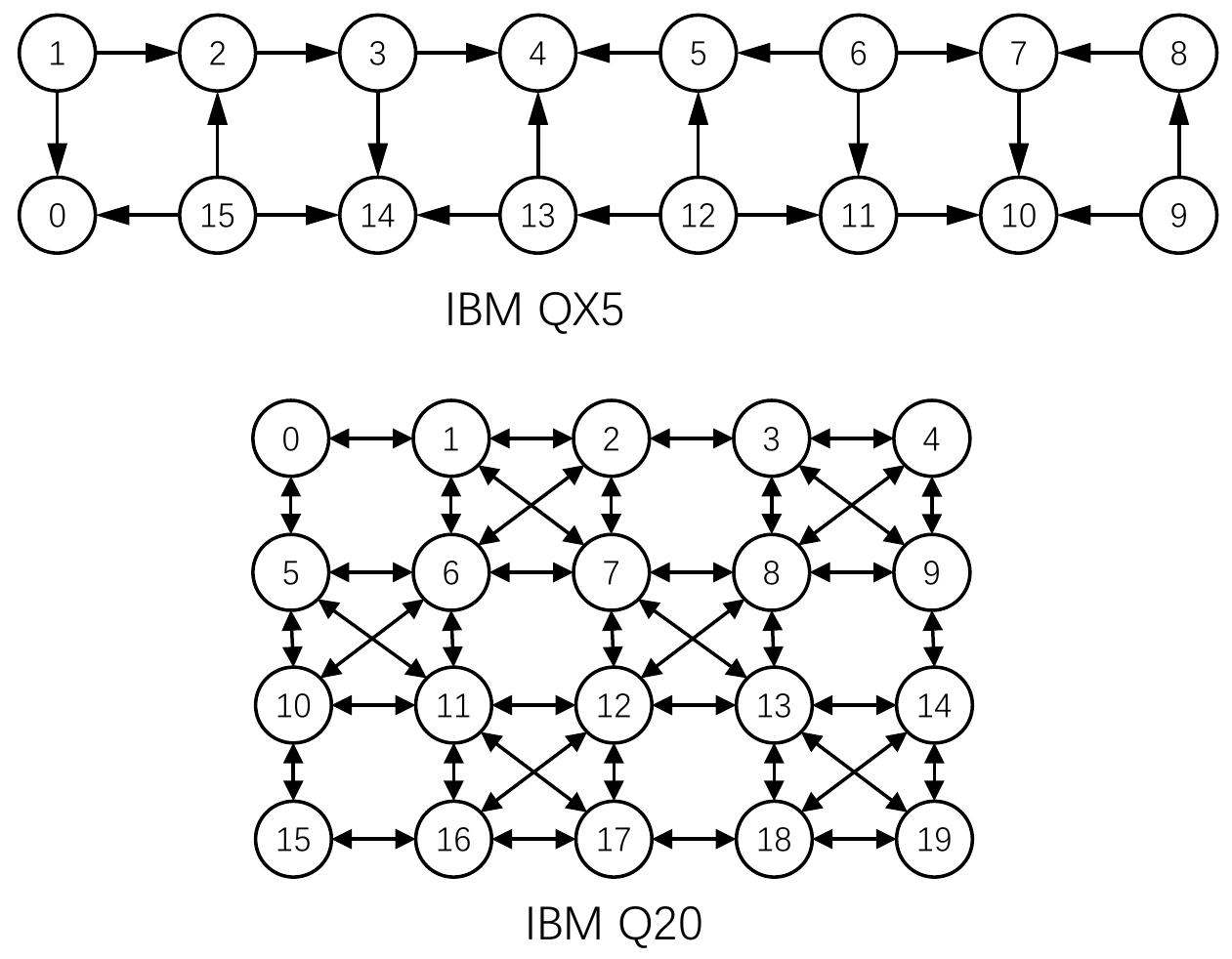}
	\caption{Two architecture graphs for IBM QX architecture.}
	\label{fig:AG}
\end{figure}

\begin{figure}[t]
	\centering
	\includegraphics[width=0.8\textwidth]{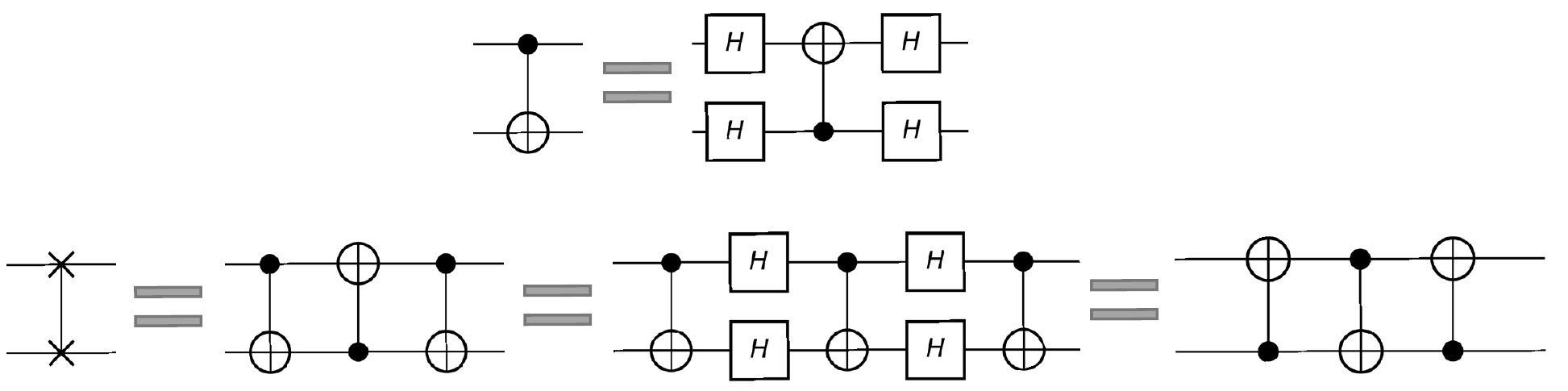}
	\caption{Some gate decomposition and transformation rules.}
	\label{fig:gate decomposition}
\end{figure}

In this paper, we only focus on the transformation process. The QPU topologies we are concerned with are IBM QX architectures QX5 and Q20 shown in Fig.~\ref{fig:AG}, but our approach is applicable to any architecture graph, including for example Rigetti 16Q Aspen-4\footnote{https://www.rigetti.com/qpu}. Notice that edges in 
IBM Q20 are bidirectional (or, undirected) and thus either node of each edge can be the control qubit of a CNOT gate. Depicted in Fig.~\ref{fig:gate decomposition} are several gate transformation rules which are quite useful in gate decomposition and circuit transformation. The top equivalence shows that we can exchange the control and target qubits of a CNOT by adding two Hadamard gates before and after it, while the bottom ones show different ways of implementing a SWAP gate in QX structures.

To simplify the presentation, we distinguish between two kinds of quantum circuits in this paper. \emph{Logical} circuits are ideal and high-level gate descriptions of quantum algorithms without considering any physical constraints imposed by QPUs. In contrast, \emph{physical} quantum circuits are low-level gate-model implementation which respect the QPU concerned. The purpose of the circuit transformation process mentioned above is then to convert a logical circuit to a physical one. Accordingly, qubits appearing in logical circuits are called logical qubits while those appearing in physical circuits are called physical ones.

\section{Quantum Circuit Transformation}
\label{sec:problem}

The main objective of quantum circuit transformation is to transform an input logical circuit to a physical one so that the constraints imposed by the QPU are satisfied. To simplify the problem, we only consider the connectivity constraints for CNOT gates as specified by the architecture graph (see Section~\ref{sec:background}). This means that single-qubit gates have no effect in the circuit transformation process, and we assume without loss of generality that the input logical circuit consists only of CNOT gates. Furthermore, a CNOT gate is simply denoted as a pair $\langle q,q'\rangle$, where $q$ is the control qubit and $q'$ is the target qubit. We call the CNOT gate $\langle q',q\rangle$ the \emph{inverse} of $\langle q,q'\rangle$.

Let $AG=(V,E)$ be the architecture graph  of a QPU, where $V$ is the set of physical qubits and $E$ the set of directed edges along which CNOT gates can be performed. Given a logical circuit $LC = (Q, C^l)$ with $|Q| \le |V|$, we need to construct a physical circuit $PC = (V, C^p)$ such that
\begin{itemize}
    \item $LC$ and $PC$ are equivalent in functionality.
    \item $C^p$ only contains CNOT gates and single qubit gates.
    \item For any CNOT gate $\langle q,q'\rangle$ in $C^p$, $( q,q') \in E$.
\end{itemize}

It is easy to find a physical circuit that satisfies the above conditions, but the real challenge is to find one with \emph{minimal} size or depth, which is NP-hard in general \cite{complexity}. 
In this paper, we modify the input logical circuit stepwise by inserting auxiliary gates like CNOT and H, as shown in Fig.~\ref{fig:gate symbol}, until the logical circuit is transformed into a physical circuit that can be executed on the QPU. To evaluate the effectiveness of quantum circuit transformation algorithms, we use the sizes of the output circuits, i.e., the total number of its elementary gates.

\subsection{Dependency Graph}
\begin{figure}
    \centering
    \includegraphics[width=0.7\textwidth]{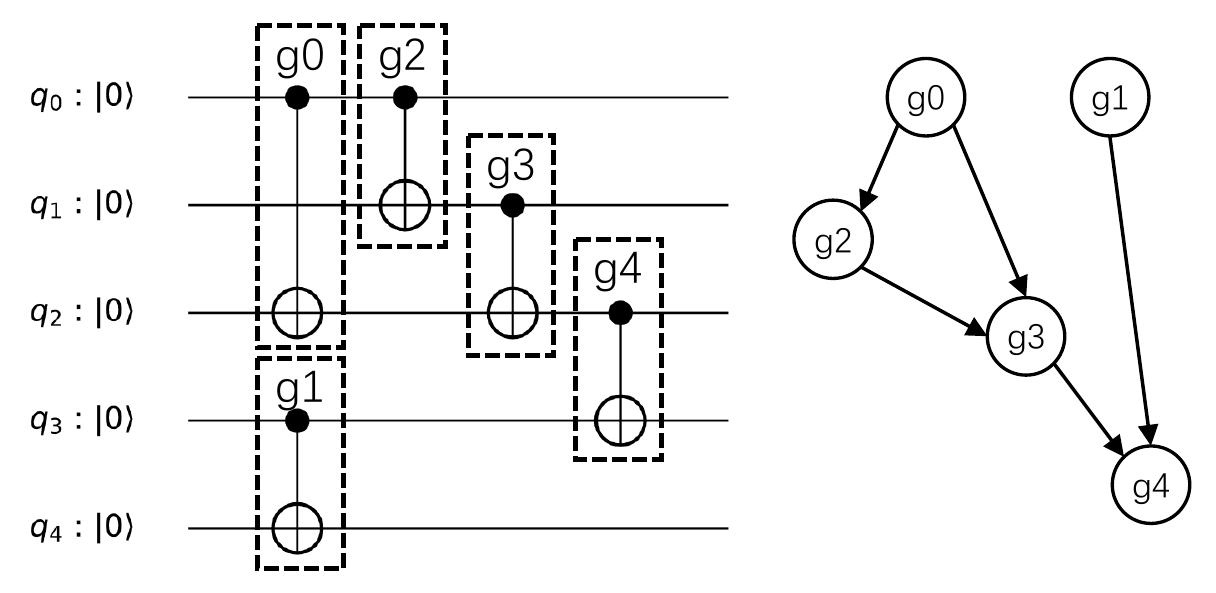}
    \caption{On the left is an example for logical quantum circuit with only CNOT gates and right DAG representing dependency order of the left circuit.}
    \label{fig:DG}
\end{figure}

CNOT gates in a logic circuit $LC = (Q,C^l)$ are not independent. We say a CNOT gate $\langle q,q'\rangle$ \emph{depends} on another $\langle p,p'\rangle$ if the latter must be executed before the former. This happens when $\langle p,p'\rangle$ is in front of $\langle q,q'\rangle$ in $C^l$ and they share a common qubit (i.e., either $p\in\{q,q'\}$ or $p'\in\{q,q'\}$), or when $\langle q,q'\rangle$ depends on a CNOT gate which depends on $\langle p,p'\rangle$.

In general, we can construct a directed acyclic graph (DAG), called the \emph{dependency graph} \cite{DG}, to characterize the dependency between gates in a logical circuit $LC$ \cite{Q20}. Each node of the dependency graph represents a gate and each directed edge the dependency relationship from one gate to another. The front layer of $LC$, denoted $\mathcal{F}(LC)$ or $\mathcal{L}_0(LC)$, consists of all gates
in $LC$ which have no parents in the dependency graph. The second layer $\mathcal{L}_1(LC)$ is then the front layer of the circuit obtained from $LC$ by deleting all gates in $\mathcal{F}(LC)$. Analogously, we can define the $k$-th layer  $\mathcal{L}_k(LC)$ of $LC$ for all $k\geq 0$.
Consider the circuit shown in Fig.~\ref{fig:DG} as an example. Initially, gates $g_0$ and $g_1$ can be applied in parallel because there are no gates before them and they are independent from each other.
Thus $\mathcal{F}(LC) = \{ g_0, g_1\}$. Then, gate $g_2$ can be executed after $g_0$, $g_3$ after $g_2$ and $g_0$, and $g_4$ after $g_3$ and $g_1$. Thus $\mathcal{L}_1(LC) = \left\{ {{g_2}} \right\}$, $\mathcal{L}_2(LC) = \left\{ {{g_3}} \right\}$, and  $\mathcal{L}_3(LC) = \left\{ {{g_4}} \right\}$.

\subsection{Qubit Mapping}
\label{sec:initial_mapping}

At each step of the circuit transformation, qubits in the logical circuit are mapped or allocated to physical qubits in the QPU  \cite{7}. Mathematically, a qubit mapping is a function $\tau$ from $Q$ to $V$ such that $\tau(q)=\tau(q')$ if and only if $q=q'$ for any $q,q'\in Q$ \cite{6}. The mapping may change at consecutive steps of the transformation which is determined by the inserted auxiliary gates.

Given a logical circuit $LC$ and a mapping $\tau$, a CNOT gate $g=\langle q,q'\rangle$ in $LC$ is said to be \emph{satisfied} by $\tau$, or $\tau$ satisfies $g$, if $(\tau(q),\tau(q'))$ is a directed edge in AG. Furthermore, $g$ is \emph{executable} by $\tau$ if it appears in the front layer of $LC$ and is satisfied by $\tau$.
In this case, we remove it from $LC$ and append a CNOT gate $\tau(g) := \langle \tau(q),\tau(q')\rangle$ to the end of the physical circuit. This process is called the \emph{execution} of $g$. 

\section{The Proposed Algorithm}
\label{sec:algorithm}

In this section, details of the proposed algorithm will be explained step by step. Let $AG = (V, E)$ be the architecture graph and $LC = (Q,C^l)$ the input logic circuit consisting only CNOT gates. The goal of the algorithm is to try to minimize the size, i.e., the total number of elementary gates, of the output physical circuit. 

We first generate an initial mapping $\tau_{ini}$ by using simulated annealing (Algorithm~\ref{alg:initial map}), and then stepwise construct the output physical circuit by adding auxiliary CNOT or Hadamard gates while processing gates in the input logic circuit. The \emph{state} of each step is described by a mapping $\tau'$ from logic qubits in $Q$ to physical qubits in $V$, the currently constructed physical circuit ${PC'}$ which obeys the constraints imposed by $AG$, and the logic circuit ${LC'}$ with gates that have not been processed. A cost function which assigns decreasing weights to gates in later layers is used to select the state of the next step.
Note that the above procedure is standard for circuit transformation, and has been adopted in~\cite{Astar}. Our algorithm distinguishes itself from the previous ones in the ways of choosing the initial mapping (Sec~\ref{sec:inimap}), the definition of the cost function, and the strategy of updating step states (Sec~\ref{sec:selnext}).

\subsection{CNOT Distance}

In graph theory, the distance from a source node $v$ to a destination node $v'$ in a directed graph $G$, written $\dist_G(v,v')$, is the minimal number of edges needed to traverse from $v$ to $v'$. Suppose AG is the architecture graph of the QPU we consider. We define the \emph{CNOT distance} from $v$ to $v'$ in AG, written $\dist_{cnot}(v,v')$, as the minimal number of auxiliary CNOT and Hadamard gates required to execute the CNOT gate $\langle v,v'\rangle$ in the QPU. Here `execute' is in the same sense as we have described in Section~\ref{sec:initial_mapping}. To execute $\langle v,v'\rangle$, we need to bring the two qubits $v$ and $v'$ close to each other by swapping and then, when they are neighbours in AG, we further check if the direction is from $v$ to $v'$ or vice versa.

For bi-directed (or, undirected) architecture graph such as that of Q20, we need only to bring $v$ close to $v'$ or vice versa, and the CNOT distance is simply computed as $\dist_{cnot}(v,v') = 3\times (\dist_{AG}(v,v')-1)$. This is because only $\dist_{AG}(v,v')-1$ swaps are required and each SWAP requires only 3 CNOT gates to implement (see Fig.~\ref{fig:gate decomposition} (top)). For directed architecture graph such as that of QX5, the situation is a little complicated, where we also need to consider the direction of the CNOT gates. We first compute all shortest paths from $v$ to $v'$ (ignoring the directions). Suppose $d=\dist_{AG}(v,v')$. If there is an undirected shortest path $\pi= \langle v_0\equiv v, v_1, ..., v_d \equiv v' \rangle$  in which $(v_i,v_{i+1})$ is a directed edge in QX5 for some $i$, then the CNOT distance is computed as $\dist_{cnot}(v,v') = 7\times (d-1)$, because a SWAP gate is decomposed into 7 elementary gates (see Fig.~\ref{fig:gate decomposition} (bottom)). Otherwise, we have $\dist_{cnot}(v,v') = 7\times (d-1) + 4$, as we need to add 2 Hadamard gates before and after to change the direction of the target CNOT \cite{Astar}.


Take QX5 as an example. Suppose the logic qubits $q$ and $q'$ are mapped to $v_3$ and $v_1$, which correspond to nodes $3$ and $1$ in Fig.~\ref{fig:AG}, respectively,  and we want to implement the CNOT gate $\langle
q,q'\rangle$, with $q$ the control qubit and $q'$ the target qubit. One solution is to add a SWAP gate between qubit $v_1$ and $v_2$ to bring $q$ one step close to $q'$, and a CNOT gate between $v_2$ and $v_3$ together with 4 additional Hadamard gates (cf. Figure~\ref{fig:gate decomposition}) to change the direction of the CNOT gate. Because a SWAP gate can be decomposed into 7 elementary gates complying with the directions in QX5, the CNOT distance from $v_3$ to $v_1$ in QX5 is 11.

For simplicity, in out algorithm we precompute the  
CNOT distance for all node pairs in AG by using, say, a breadth-first search algorithm.

\subsection{Initial Mapping}\label{sec:inimap}

The selection of a good initial mapping has a significant impact on the quality of the final physical circuit \cite{initialmap,Q20,Astar}.  Intuitively, we would like to find an initial mapping that `fits' most gates such that fewer SWAP gates are required in the circuit transformation process.

To this end, we define the \emph{gate cost} of a CNOT gate $g=\langle q,q'\rangle$ under a mapping $\tau: Q \to V$ as 
\begin{equation}
\cost_{gate}(g, \tau) = {\dist_{cnot}}(\tau(q),\tau(q')).
\end{equation}
Our ideal initial mapping $\tau_{ini}^*$ is then given by  
\begin{equation}
\label{eq:goal for initial map}
\tau_{ini}^* = \argmin_\tau \left\{ {\sum\limits_{g \in {C^*}} {\cost_{gate}(g,\tau)} } \right\}
\end{equation}
where $C^*$ is a selected subset of the logical circuit $LC$. Here we use $C^*$ instead of $C^l$ to calculate the initial mapping. This is because taking into account all gates in $C^l$ would bring further overhead and be unnecessary because gates in the tail of the circuit would have little impact on the initial mapping.


\begin{figure}
    \centering
    \includegraphics[width=0.55\textwidth]{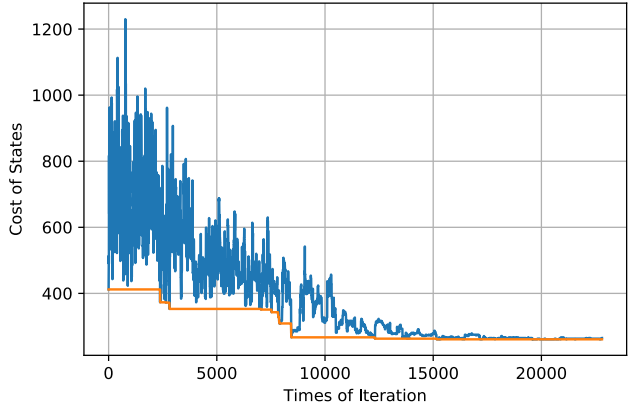}
    \caption{Convergence of the simulated annealing algorithm on circuit adr4-197  and IBM Q20, where the blue and orange lines represent the cost of accepted states and existing best states, respectively. We set empirically $T_{\max}=100$, $T_{\min}=1$, $\Delta=0.98$, and $R=100$.}
    \label{fig:SA_convergence}
\end{figure}

Simulated annealing (SA), inspired by the annealing process in metallurgy \cite{annealing}, is designed for approximating the global optimum of a given cost function. The algorithm tries to find the best state in the search space. In each trial for searching a better state, the algorithm generates a new state based on the previous one, calculates its cost and compares it with the previous one and decides whether this new state should be accepted. To escape from local optima, the algorithm accepts the new generated state with a certain probability even if its cost is worse than the previous one. The acceptance probability is decided by the current temperature which declines during the search process until the minimum value is reached.

We propose an efficient simulated annealing based algorithm (Algorithm~\ref{alg:initial map}) to find a good approximation of $\tau_{ini}^*$, where $T_{\max}$, $T_{\min}$, $\Delta$ and $R$ are, respectively, the starting temperature, the minimum temperature, the decline coefficient for the temperature and the repeated times for one temperature. Fig.~\ref{fig:SA_convergence} shows convergence of the simulated annealing process on a real quantum circuit adr4-197. Note that the cost of states converges after sufficient iterations, showing that the temperature is low enough. The fluctuation of the cost of accepted states is caused by the above mentioned acceptance probability for worse states.


\begin{figure}[!ht]
	
	\begin{algorithm}[H]
		\SetKwData{Left}{left}\SetKwData{This}{this}\SetKwData{Up}{up}
		\SetKwFunction{Union}{Union}\SetKwFunction{FindCompress}{FindCompress}
		\SetKwInOut{Input}{input}\SetKwInOut{Output}{output}
		\Input{A set $C^*$ for considered gates in a logical quantum circuit.}
		\Output{An approximation of the optimal initial mapping given in Eq.(\ref{eq:goal for initial map}).}
		\caption{Simulated annealing for computing the initial mapping}\label{alg:initial map}
		
		\Begin{	
			
			Initialize parameters $T_{\max}$, $T_{\min}$, $\Delta$, $R$, and an arbitrary mapping $\tau$\;
			$T \leftarrow T_{\max}$, $bcost \leftarrow \infty$, $cost \leftarrow \infty$\;
			\While{$T \ge T_{\min}$}
			{
				$i \leftarrow 1$\;
				\While{$i \leq R$}
				{
					$i \leftarrow i+1$\;
					Change mapping $\tau$ randomly to generate a new mapping $\tau_{new}$\;
					$ncost = {\sum_{g \in {C^*}} {{\cost_{gate}}\left( {g,\tau_{new}} \right)}}$\;
					\If{$ncost <  bcost$}
					{
						$bcost \leftarrow ncost$\;
						$\tau_{ini} \leftarrow \tau_{new}$\;
					}
				\eIf{$ncost <  cost$}
				{
					$cost \leftarrow ncost$\;
					$\tau \leftarrow \tau_{new}$\;
				}
				{
					$cost \leftarrow ncost$ \mbox{and} $\tau \leftarrow \tau_{new}$ \mbox{with prob.} $\exp(\frac{cost - ncost}{T})$\;
				}
				
			}
			
			$T \leftarrow \Delta \times T$;
			}
			\Return{$\tau_{ini}$}
		}
	\end{algorithm}
\end{figure}

In the following two subsections, we describe in detail how to generate all possible child and grandchild states and how the cost of a child or grandchild state is calculated. We will illustrate the construction by using the  quantum circuit shown in Fig.~\ref{fig:alu-v0} and the test architecture graph $AG_{\sf test}$ in Fig.~\ref{fig:agtest}.

\subsection{Heuristic search with look-ahead}
\label{sec:selnext}

We have shown in the previous section how to construct the initial mapping $\tau_{ini}$ for our circuit transformation algorithm, thus obtaining the state $s^{0} := (\tau_{ini}, PC_0, LC_0)$ for the first step, where $PC_0$ and $LC_0$ are respectively the physical and logic circuits after executing all gates in $LC$ which are executable in $\tau_{ini}$. Suppose we are in state $s^{i} := (\tau_i, PC_i, LC_i)$ at the $i$-th step for $i\geq 0$. This section is devoted to the strategy of choosing $s^{i+1}$ for the $i+1$-th step. Obviously, depending on the different ways of adding auxiliary CNOT and H gates, there are multiple child states of $s^i$ to choose from. One natural way is to select the one with the minimal cost. This surely gives a fine method for extending $s^i$, but (as shown in Fig.~\ref{fig:depth}) the sizes of the output physical circuits are not always desirable. In this paper, we propose a novel way to select the next state: we look one level ahead to calculate the costs of all \emph{grandchild} states of $s^i$, and choose the child of $s^i$ which has a child (thus a grandchild of $s^i$) with the minimum cost among all grandchildren of $s^i$.

To this end, we have to specify for a given state $s^i := (\tau_i, PC_i, LC_i)$, (1) how to extend $s^i$ to get all its children and grandchildren, and (2) how to define the costs of its grandchildren.
We are going to elaborate these two points one by one in the following. 

\textbf{Extend $s^i$}. There are two natural ways to {extend} $s^i$.
\begin{itemize}
	\item Way 1: Apply on $\tau_i$ a swap operation represented as an edge in $AG$ one of whose end nodes is the image under $\tau_i$ of some qubit appearing in a gate in the front layer of ${LC}_i$, and obtain a new mapping $\tau_i'$. Accordingly, we extend ${PC}_i$ with the CNOT + H implementation of the SWAP gate corresponding to this swap operation. Then we execute recursively all gates in $LC_i$ (not only those in the front layer, but also those executable when their precedents have already been executed by $\tau_i'$) which are executable in $\tau_i'$.
	The resultant state is then a child of $s^i$. 
	
	\item Way 2 only applies when AG is directed and there is a CNOT gate $\langle q,q'\rangle$ in the front layer of ${LC}_i$ which is inversely executable, i.e. its inverse gate $\langle q',q\rangle$ is executable, in $\tau_i$. In this case, we add 4 Hadamard gates to change the direction of $\langle q,q'\rangle$  (cf. Figure~3 (top)), extend $PC_i$ with all these 5 gates, and delete $\langle q,q'\rangle$ from $LC_i$. Again, we execute recursively all gates in $LC_i$ which are executable in $\tau_i$ to get a child of $s^i$. 	
\end{itemize}

Finally, for each child of $s^i$, we extend one level further to get its grandchildren.
We denote by $\{s^i_j : j\in J\}$ and $\{s^i_{j,k} : j\in J, k\in K\}$ the set of children and grandchildren of $s^i$, respectively. 

\begin{figure}
	\centering
	\includegraphics[width=0.95\textwidth]{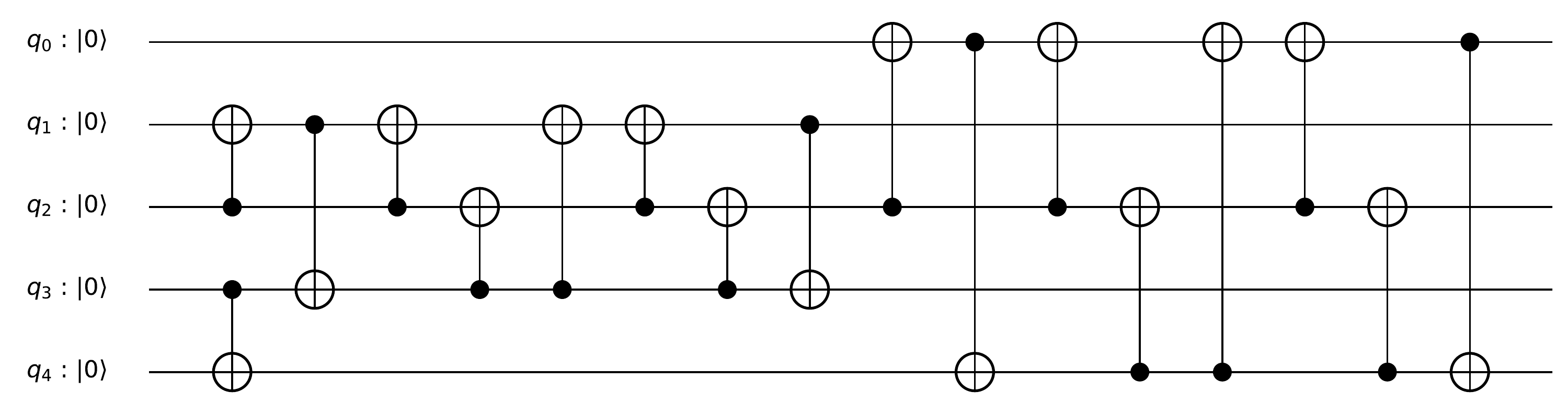}
	\caption{The quantum circuit alu-v0$\_$27 with all single qubit gates removed.}
	\label{fig:alu-v0}
\end{figure}

\begin{figure}
	\centering
	\includegraphics[width=0.25\textwidth]{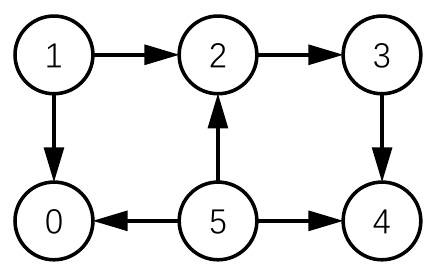}
	\caption{A test architecture graph $AG_{\sf test}$.}
	\label{fig:agtest}
\end{figure}


\begin{example}
	\label{ex:alu}
	We consider the quantum circuit shown in Fig.~\ref{fig:alu-v0} and the test architecture graph $AG_{\sf test}$ in Fig.~\ref{fig:agtest}. Applying Alg.~\ref{alg:initial map} we get the initial mapping $\tau_{ini}:Q \to V$ which maps $q_i$ to $v_i$ for each $0\leq i\leq 4$. For convenience, we write such a mapping as a list of length 5. For example, $\tau_{ini} =[0,1,2,3,4]$. Note that the front layer contains two gates, viz., $\langle q_2,q_1\rangle$ and $\langle q_3,q_4\rangle$. As the latter is directly executable by $\tau_{ini}$, the initial state $s = (\tau_{ini}, PC_0 := \{\langle q_3,q_4\rangle\}, LC_0 := LC \backslash \{\langle q_3,q_4\rangle\})$, where $LC$ is the circuit shown in Fig.~\ref{fig:alu-v0}.
	
	 We next show how to construct the child states of $s$. Note that there is only one gate, viz. $\langle q_2,q_1\rangle$, in the first layer of $LC_0$, and $\tau_{ini}$ maps $q_1$ and $q_2$ to, respectively, $v_1$ and $v_2$. Only 4 edges, $(1,0), (1,2),(2,3)$ and $(5,2)$, in $AG_{\sf test}$ (see Fig.~\ref{fig:agtest}) are relevant. For each of them, we obtain a corresponding swap operation and a corresponding child state. Since $AG_{\sf test}$ is directed and $\tau_{ini}$ can execute $\langle q_1,q_2\rangle$, another child state can be obtained by using Way 2. Therefore, $s$ has in total 5 child states and Table~\ref{tab:childstates_of_0} gives the mappings and physical circuits of these child states as well as the corresponding operations. Similarly, we also construct the grandchild states of $s$, also shown in Table~\ref{tab:childstates_of_0}. Here in the `Operation' column, we use an edge in $AG_{\sf test}$ to denote the corresponding swap operation on mappings, and a CNOT gate to denote the operation of changing its direction.  
\end{example}

\begin{table}[t]
	\centering
	\begin{tabular}{c|ccccc}
		Child State & Mapping & Newly added gates in $PC_s$ & Operation & $\cost_g$ & $\cost_h$ \\ \hline
		$s_{0}$ & [1,0,2,3,4] & $\{  \swap{0}{1}\}$ & (1,0) &7 &208.2  \\
		$s_{1}$ & [1,0,2,3,4] & $\{ \swap{1}{2}, \langle 1,2\rangle, \langle 2,3\rangle, \langle 1,2\rangle\}$ & (1,2) &7 &167.2  \\
		$s_{2}$ & [0,1,3,2,4] & $\{ \swap{2}{3}\}$ & (2,3) &7 &199.0  \\
		$s_{3}$ & [0,1,5,3,4] & $\{ \swap{5}{2}\}$ & (5,2) &7 &203.0  \\
		$s_{4}$ & [0,1,2,3,4] & 
		$\{ \overleftarrow{\langle 1,2\rangle}\}$
		& $\langle q_2,q_1\rangle$ &4 &188.8  \\
		$s_{0,0}$ & [0,1,2,3,4] &$\{ \swap{0}{1} \}$& (0,1) &14&195.8  \\
		$s_{0,1}$ & [1,5,2,3,4] &$\{ \swap{0}{5} \}$& (0,5) &14&195.8  \\
		$s_{0,2}$ & [2,0,1,3,4] &$\{ \swap{1}{2}, \langle 1,0\rangle \}$& (1,2) &14&199.2  \\
		$s_{0,3}$ & [1,0,3,2,4] &$\{ \swap{2}{3} \}$& (2,3) &14&211.4  \\
		$s_{0,4}$ & [1,0,5,3,4] &$\{ \swap{2}{5}, \langle 5,0\rangle \}$& (2,5) &14&196.0  \\
		$s_{1,0}$&[1,2,0,3,4]&$\{ \swap{0}{1} \}$&(0,1)&14&177.6  \\
		$s_{1,1}$&[0,1,2,3,4]&$\{ \swap{1}{2} \}$&(1,2)&14&166.2  \\
		$s_{1,2}$&[0,3,1,2,4]&$\{ \swap{2}{3} \}$&(2,3)&14&157.6  \\
		$s_{1,3}$&[0,2,1,4,3]&$\{ \swap{3}{4} \}$&(3,4)&14&175.0  \\
		$s_{2,0}$&[1,0,3,2,4]&$\{ \swap{0}{1} \}$&(0,1)&14&211.4  \\
		$s_{2,1}$&[0,2,3,1,4]&$\{ \swap{1}{2} \}$&(1,2)&14&194.6  \\
		$s_{2,2}$&[0,1,2,3,4]&$\{ \swap{2}{3} \}$&(2,3)&14&195.8  \\
		$s_{2,3}$&[0,1,4,2,3]&$\{ \swap{3}{4} \}$&(3,4)&14&208.6  \\
		$s_{3,0}$&[1,0,5,3,4]&$\{ \swap{0}{1}, \langle 5,0\rangle \}$&(0,1)&14&196.0  \\
		$s_{3,1}$&[5,1,0,3,4]&$\{ \swap{0}{5} \}$&(0,5)&14&201.8  \\
		$s_{3,2}$&[0,2,5,3,4]&$\{ \swap{1}{2}, \langle 5,2\rangle, \langle 2,3\rangle, \langle 5,2\rangle \}$&(1,2)&14&160.8  \\
		$s_{3,3}$&[0,1,2,3,4]&$\{ \swap{2}{5} \}$&(2,5)&14&195.8  \\
		$s_{3,4}$&[0,1,4,3,5]&$\{ \swap{4}{5} \}$&(4,5)&14&211.4  \\
		$s_{4,0}$&[1,0,2,3,4]&$\{ \swap{0}{1} \}$&(0,1)&11&200.6  \\
		$s_{4,1}$&[0,2,1,3,4]&$\{ \swap{1}{2}, \langle 2,3\rangle, \langle 1,2\rangle \}$&(1,2)&11&167.2  \\
		$s_{4,2}$&[0,1,3,2,4]&$\{ \swap{2}{3}, \langle 1,2\rangle \}$&(2,3)&11&177.6  \\
		$s_{4,3}$&[0,1,2,4,3]&$\{ \swap{3}{4} \}$&(3,4)&11&200.0  \\
	\end{tabular}
	\caption{The child and grandchild states of $s = (\tau_{ini}, PC_0 := \{\langle q_3,q_4\rangle\}, LC_0 := LC \backslash \{\langle q_3,q_4\rangle\})$ in Example~\ref{ex:alu}, where $i \leftrightarrows j$ denotes the swap operation of $i$ and $j$ and $\mathop {\left\langle {\left. {i,j} \right\rangle } \right.}\limits^ \leftarrow $ denotes the operation that changes the direction of the CNOT gate $\langle i,j \rangle$.}
	\label{tab:childstates_of_0}
\end{table}

\textbf{Evaluate the grandchildren of $s^i$}. 
The cost of a grandchild $s^i_{j,k}$ of $s^i$ consists of two parts: 
the first part, $\cost_g(s^i_{j,k})$, is the number of auxiliary CNOT and Hadamard gates added during the evolution from $s^i$ to $s^i_{j,k}$, and the second part, $\cost_h(s^i_{j,k})$, is an estimated cost for completing the remaining gates in the logic circuit of $s^i_{j,k}$.

The first part depends on the different ways of extending $s^i$ and $s^i_j$ to obtain $s^i_{j,k}$. 
If $AG$ is undirected, then only Way 1 is available for the extensions and 3 CNOT gates suffice to implement the required swap operation on the mapping. Thus $\cost_g(s^i_{j,k}) = 6$.
 Otherwise, 7 gates (3 CNOTs and 4 Hadamard shown in Fig.~\ref{fig:gate decomposition}) for Way 1 and 4 Hadamard for Way 2 are needed. Thus $\cost_g(s^i_{j,k})$ can be 14, 11, or 8. 
 Consider the state $s$ in Example~\ref{ex:alu}. From Table~\ref{tab:childstates_of_0} we can see that each grandchild state of $s$ has $\cost_g$ 14 or  11.

For the second part of the cost, we employ a look-ahead mechanism first demonstrated in \cite{Q20}. 
Given a generic state $s = (\tau_s, PC_s, LC_s)$, we partition the gates in ${LC}_{s}$ into different layers according to its dependency graph. Denote by $L_k$, $k\geq 0$, these layers such that $L_0$ is the front layer. Then the heuristically estimated cost of $s$ is defined as 
\begin{equation}
\label{eq:5}
\cost_h(s) = \sum_{k=0}^\ell w_k \left(\sum_{g \in L_k} \cost_{gate}(g,\tau_s) \right) + w_{s}\times (d-1) \times N_{swap} \times N_s, 
\end{equation}
where $d$ is the diameter of the architecture graph, $N_{swap}$ the number of elementary gates needed to compose a SWAP gate, and $N_{s}$ is the number of gates in $LC_s$.
The parameters
 $\ell>0$, $w_k$ $(0\leq k \leq \ell)$ and $w_s$ are taken empirically but normally we assume $1=w_0 \geq w_1 \geq \cdots \geq w_l \geq w_s \geq 0 $. This reflects the intuition that the closer a gate is from the front layer of the circuit, the more it contributes to the total cost of executing the whole circuit, as subsequent dependent gates will not be executable unless it has been processed.   
 Table~\ref{tab:childstates_of_0} gives the heuristic costs for all child and grandchild states of $s$ in Example~\ref{ex:alu}, where we take $\ell=3$, $w_1= 1$, $w_2= 0.8$, $w_3= 0.6$, $w_s=0.4$. Note that the diameter of QX5 is $8$ and each SWAP gate is composed by 7 elementary  gates in QX5. Thus $d=8$ and $N_{swap}=7$.

Finally, the total cost of a grandchild $s^i_{j,k}$ of $s^i$ is computed as
\begin{equation}
\label{eq:cost}
\cost(s^i_{j,k}) = \cost_g(s^i_{j,k}) + \cost_h(s^i_{j,k}).
\end{equation}
Suppose $s^i_{j^*,k^*}$ is a grandchild state with the minimum cost. Then $s^i_{j^*}$ is selected as the state for the $i+1$-th step; that is, we let $s^{i+1} = s^i_{j^*}$. 
For the state $s$ in Example~\ref{ex:alu}, we can see from Table~\ref{tab:childstates_of_0} that $s_{1,2}$ is the grandchild state with minimum cost. Thus we select its parent $s_1$ as our next state. Note that $s_1$ happens to be the child state which also has the minimum cost among all child states of $s$. In general, this coincidence does not hold.
The whole algorithm for circuit transformation is shown in Algorithm~\ref{alg:main}.

\begin{figure}[!ht]
	
	\begin{algorithm}[H]
		\SetKwData{Left}{left}\SetKwData{This}{this}\SetKwData{Up}{up}
		\SetKwFunction{Union}{Union}\SetKwFunction{FindCompress}{FindCompress}
		\SetKwInOut{Input}{input}\SetKwInOut{Output}{output}
		\Input{A logic circuit $LC = (Q, C^l)$, an initial mapping $\tau$ constructed by Algorithm~\ref{alg:initial map} and an architecture graph $AG = (V, E)$ with $|Q|\leq |V|$.}
		\Output{A physical circuit $(V, C^p)$ which satisfies $AG$ and is equivalent to $LC$.}
		\caption{Circuit transformation with look-ahead}\label{alg:main}
		
		\Begin{	
			
			$(PC, LC) \leftarrow \mbox{Execute}(\tau, PC, LC)$\;
			\While{$LC \neq \emptyset$}
			{
				$L\leftarrow \mathcal{F}(LC)$, \mbox{the first layer of $LC$}\;
				$Cld\leftarrow \emptyset$\;
				\For{$e\in E$ which touches some gate in $L$ under $\tau$}
				{
					$\tau' \leftarrow \mbox{swap}_e \circ \tau$\;
					$PC' \leftarrow PC'$ by adding (the $CNOT$ + $H$ implementation of) a $\mbox{SWAP}$ gate corresponding to swap$_e$\;
					
					$(PC', LC') \leftarrow \mbox{Execute}(\tau', PC', LC)$\;
					$gcost \leftarrow $ \mbox{3 if $e^{-1}\in E$ and 7 otherwise}\;
					$Cld \leftarrow Cld\cup \{(\tau', PC', LC', gcost)\}$\;			
				}
				
				\For{$g\in L$ which is inversely executable by $\tau$}
				{
					$PC' \leftarrow$ $PC'$ by adding $\tau(g)$ complemented by four $H$ gates before and after it\;
					$LC' \leftarrow$ $LC'$ by deleting $g$\;						
					$(PC', LC') \leftarrow \mbox{Execute}(\tau, PC', LC')$\;
					$Cld \leftarrow Cld\cup \{(\tau, PC', LC', 4)\}$\;
				}
			
				$mCost \leftarrow \infty$\;
				\For{$(\tau', PC', LC', gcost)\in Cld$}
				{	
					$cost\leftarrow \mbox{minChildHcost}(\tau', LC')$\;
					\If{$cost + gcost < mCost$}
						{
							$mCost \leftarrow cost + gcost$\;
							$(\tau, PC, LC) \leftarrow (\tau', PC', LC')$\;
						}
				}
			}
			\Return{$PC$}
		}
	\end{algorithm}
\end{figure}

\SetKwRepeat{Do}{do}{while}
\begin{figure}[!ht]
	
	\begin{procedure}[H]
		\SetKwData{Left}{left}\SetKwData{This}{this}\SetKwData{Up}{up}
		\SetKwFunction{Union}{Union}\SetKwFunction{FindCompress}{FindCompress}
		\SetKwInOut{Input}{input}\SetKwInOut{Output}{output}
		
		\Input{A mapping $\tau : Q\rightarrow V$, a physical circuit $PC$, and a logic circuit $LC$.}
		\Output{A pair $(PC', LC')$ obtained by executing as many as possible gates which satisfy $\tau$.}
		\caption{Execute($\tau, PC, LC$)}
		\Begin{
			$PC' \leftarrow PC$; $LC' \leftarrow LC$\;
			\Do{$EL\neq \emptyset$}
			{

				$EL \leftarrow \{g\in  \mathcal{F}(LC’) : g\mbox{ is executable by }\tau\}$\;	
			
				\For{$g\in EL$}
				{
					$PC' \leftarrow$ $PC'$ by adding $\tau(g)$\;
					$LC' \leftarrow$ $LC'$ by deleting $g$\;
				}
			}
		\Return{$(PC', LC')$}
			}
	\end{procedure}
\end{figure}

\begin{figure}[!ht]
	
	\begin{procedure}[H]
		\SetKwData{Left}{left}\SetKwData{This}{this}\SetKwData{Up}{up}
		\SetKwFunction{Union}{Union}\SetKwFunction{FindCompress}{FindCompress}
		\SetKwInOut{Input}{input}\SetKwInOut{Output}{output}
		
		\Input{A mapping $\tau : Q\rightarrow V$ and a logic circuit $LC$.}
		\Output{The minimal cost of all children of $\tau$.}
		\caption{minChildHcost($\tau, LC$)}
		\Begin{
			$L\leftarrow \mathcal{F}(LC)$\;
			$mCost \leftarrow \infty$\;
			\For{$e\in E$ which touches some gate in $L$ under $\tau$}
			{
				$\tau' \leftarrow \mbox{swap}_e \circ \tau$\;
				$(PC', LC') \leftarrow \mbox{Execute}(\tau',\emptyset, LC)$\;
				$gcost \leftarrow $ \mbox{3 if $e^{-1}\in E$ and 7 otherwise}\;
				$hcost\leftarrow hcost(\tau', LC')$ according to Eq.\eqref{eq:5}\;
				\If{$hcost + gcost < mCost$}
				{
					$mCost \leftarrow hcost + gcost$\;
				}			
			}
			
			\For{$g\in L$ which is inversely executable by $\tau$}
			{
				$LC' \leftarrow$ $LC$ by deleting $g$\;
				$(PC', LC') \leftarrow \mbox{Execute}(\tau, \emptyset, LC)$\;	
				$hcost\leftarrow hcost(\tau, LC')$ according to Eq.\eqref{eq:5}\;		
				\If{$hcost + 4 < mCost$}
				{
					$mCost \leftarrow hcost + 4$\;
				}
			}
		
			\Return{mCost}\;
		}
	\end{procedure}
\end{figure}

\subsection{Fallback via remote CNOT}
During the search process, there is a small possibility that our algorithm does not halt. This happens when a child state with better cost may be good for gates in look-ahead layers but increases the distances of gates in the front layer. To address this problem, a fallback mechanism is introduced to ensure that the program terminates in reasonable time.

\begin{figure}
    \centering
    \includegraphics[width=0.4\textwidth]{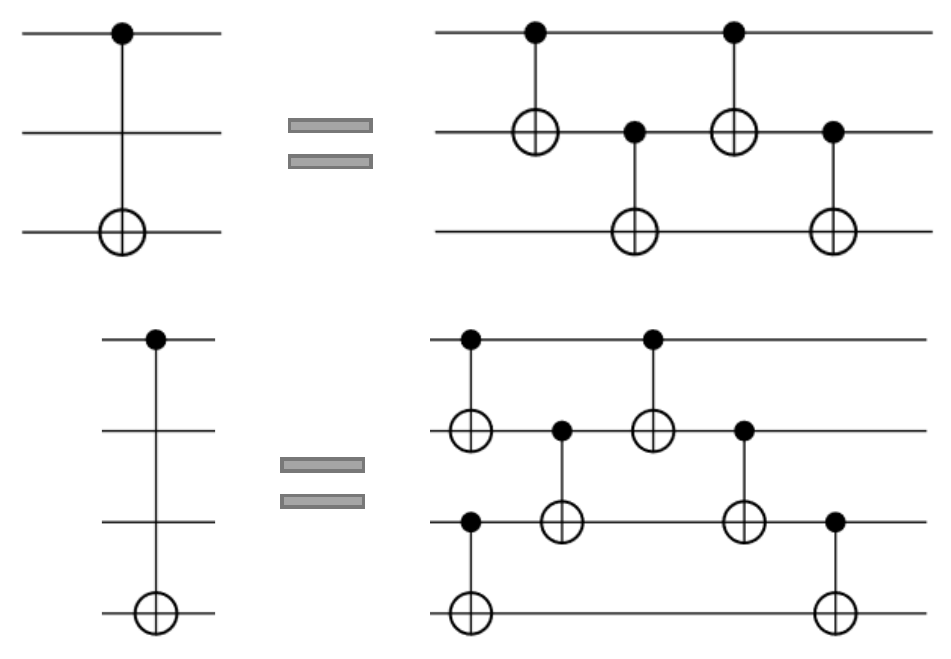}
    \caption{Schematic for remote CNOT operations with 2 and 3 hops. Generalized form can be found in \cite{Ud1}.}
    \label{fig:remoteCNOT}
\end{figure}


A direct way for fallback is to select a gate $\langle q,q'\rangle$ in the front layer and then choose a SWAP operation that will reduce the shortest path between the two corresponding nodes $v,v'$ with $\tau_s(q)=v$ and  $\tau_s(q')=v'$ in the architecture graph \cite{6}, where $s$ denotes the current state. However, this method may change the mapping that the algorithm may want to keep as it is preferred by look-ahead layers. To protect the preferred mapping, remote CNOT operations \cite{Rodney}, which are depicted in Fig.~\ref{fig:remoteCNOT}, are introduced in the fallback. After imposing remote CNOT gates, the circuit has the same functionality while preserving the current mapping. The fallback is activated when no gates are removed from ${LC}_s$ after a certain prefixed number of rounds.

\subsection{Complexity of the Search Process}

In each layer, there are at most $| Q |/2$ gates, where $Q$ is the set of qubits in the input logic circuit. Thus, the time complexity of computing the cost (cf. Eq.\eqref{eq:cost}) of any state is $O(\ell \cdot |Q|)$, where $\ell$ is the prefixed small number of layers we select for Eq.\eqref{eq:5}. For our evaluation (see Section~\ref{sec:benchmark}), we take $\ell=3$ for all circuits.


By construction, each state $s$ has at most  $|E|+|Q|/2$ child states, where $E$ is the set of edges in the architecture graph, or, equivalently, the number of possible SWAP operations that can be added to the circuit and $|Q|/2$ is the number of CNOT gates in the front layer of the current logic circuit that can be applied by adding four extra Hadamard gates to change the direction.

Suppose the input circuit contains $m$ CNOT gates. 
If we activate the fallback when no gates are removed from ${LC}_s$ after $K$ rounds, then the search procedure has at most 
$K\times m$
states. This is because each activation of the fallback will execute a selected gate due to the use of remote CNOT. 
Therefore, the overall time complexity of the search is $O(\ell\cdot |Q| \cdot (|E|+|Q|/2)^2 \cdot  m \cdot K)$. Because $|Q|\leq |V|$ and $|E| \leq |V|\cdot (|V|-1)/2 $, it is bounded by $O(|V|^4 \cdot |Q| \cdot \ell \cdot m \cdot K)$.

For the space complexity, in each state $s$, we maintain a depth-$2$ search tree rooted at $s$. Thus the space complexity of the algorithm is bounded by $O((|E|+|Q|/2)^{2})$, i.e., $O(|V|^4)$.

\subsection{Optimization}
\label{sec:modification}
In Algorithm \ref{alg:main}, the search space grows exponentially if the depth of look-ahead is increased. Therefore, a pruning mechanism is introduced to reduce the size of the search space while preserving the quality of the output physical circuit. More specifically, a child state $s^i_j$ of $s^i$ will be removed if 
both $\cost_{h}'(s^i_{j}) > \cost_{h}'(s^i)$ and $\cost_h(s^i_j) - \cost_{h}'(s^i_{j}) > \cost_h(s^i)  - \cost_{h}'(s^i)$, where $\cost'_{h}(s) = w\times (d-1) \times N_{swap} \times N_ss$ as defined in Eq.(\ref{eq:5}). In Example.~\ref{ex:alu}, states $s_{1,0}$ will be pruned. This is because $\cost_{h}'(s)=145.6$, $\cost_{h}(s)=167.2$, $\cost_{h}'(s_0)=145.6$, $\cost_{h}(s_0)=177.6$ and $\cost_{h}(s_0)>\cost_{h}(s)$, $\cost_h(s_0) - \cost_{h}'(s_{0})> \cost_h(s)  - \cost_{h}'(s) $.

\begin{figure}
    \centering
    \includegraphics[width=0.6\textwidth]{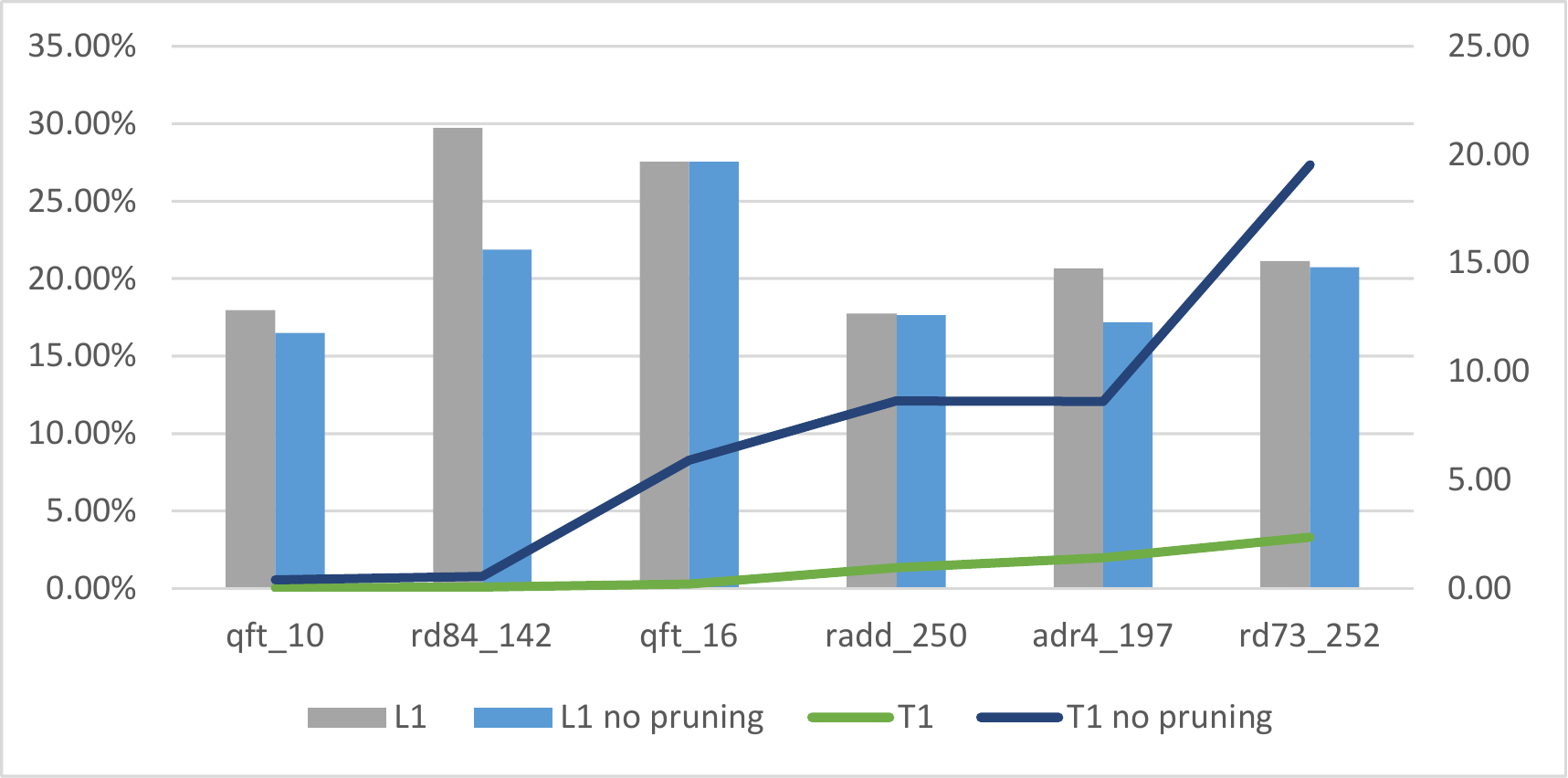}
    \caption{Effectiveness of the pruning mechanism obtained by running exemplary circuits on IBM Q20, where the gray and blue bars are ratios of the number of added gates to that of original gates for the proposed algorithm with and without pruning while lines correspond to the time consumption (seconds) for the search process.}\label{fig:pruning}
\end{figure}

From Fig.~\ref{fig:pruning} we can see that the pruning mechanism has limited influence on the sizes of the output circuits while the time consumption is reduced by a large amount.

\section{Programming and Benchmarks}
\label{sec:benchmark}

To evaluate our approach, we compare it with previous algorithms proposed for the same purpose in the literature~\cite{Astar,Q20,11}. We use Python as our programming language and IBM Qiskit \cite{qiskit} as auxiliary environment. The code can be found in GitHub\footnote{https://github.com/BensonZhou1991/circuittransform/}. All experiments are conducted in a laptop with i7-8750H CPU and 16GB memory. The results are reported in Tables~\ref{tab:SA}, ~\ref{tab:QX5} and \ref{tab:Q20}, in which the `Comparison' column shows the improvement of our algorithm over previous ones in terms of the numbers of auxiliary gates added. Specifically, let $n_{comp}$ and $n_{ours}$ be the numbers of gates added by the compared algorithm and by ours, respectively. Then the improvement ratio is defined as $(n_{comp} - n_{ours})/n_{comp}$.

Table~\ref{tab:SA} demonstrates the superiority of the initial mapping output by our simulated annealing algorithm (Alg.~\ref{alg:initial map}) compared to the naive initial mapping which maps $q_i$ to $v_i$ for all $i$ in Q20. The improvement is consistent and often above  $30\%$.

For the heuristic search, we compare our algorithm to the ones introduced in \cite{Astar} and \cite{Q20}, which are, respectively, the state-of-the-art algorithms for IBM QX5 and Q20. We set the number of look-ahead layers as $\ell=3$, and the weight parameters $w_1=1$, $w_2=0.8$, $w_3=0.6$, $w_4=0.4 \times \left( {{D_{AG}} - 1} \right) \times N_{swap}$ in Eq.\eqref{eq:5}, where $D_{AG}$ is the diameter of the architecture graph and $N_{swap}$ the number of elementary gates needed to compose a SWAP gate. The threshold number $K$ for activating fallback is set to be $0.5 \times D_{AG}$.

\begin{table}
    \centering
    \begin{tabular}{ccccc}
    \hline
    \begin{tabular}[c]{@{}c@{}}Circuit\\   Name\end{tabular} & \begin{tabular}[c]{@{}c@{}}Original\\ Gates\end{tabular} & \begin{tabular}[c]{@{}c@{}}Naive\\ Mapping\end{tabular} & \begin{tabular}[c]{@{}c@{}}Simulated\\ Annealing\end{tabular} & Improvement \\
    \hline
    4mod5-v1\_22                                             & 21                                                       & 30                                                      & 21                                                            & 100.00\%   \\
    mod5mils\_65                                             & 35                                                       & 56                                                      & 35                                                            & 100.00\%   \\
    alu-v0\_27                                               & 36                                                       & 51                                                      & 42                                                            & 60.00\%    \\
    decod24-v2 43                                            & 52                                                       & 85                                                      & 52                                                            & 100.00\%   \\
    4gt13\_92                                                & 66                                                       & 120                                                     & 66                                                            & 100.00\%   \\
    ising\_model\_10                                         & 480                                                      & 528                                                     & 480                                                           & 100.00\%   \\
    ising\_model\_13                                         & 633                                                      & 687                                                     & 633                                                           & 100.00\%   \\
    ising\_model\_16                                         & 786                                                      & 882                                                     & 786                                                           & 100.00\%   \\
    qft\_10                                                  & 200                                                      & 290                                                     & 236                                                           & 60.00\%    \\
    qft\_16                                                  & 512                                                      & 740                                                     & 647                                                           & 40.79\%    \\
    rd84\_142                                                & 343                                                      & 490                                                     & 445                                                           & 30.61\%    \\
    adr4\_197                                                & 3439                                                     & 4483                                                    & 4150                                                          & 31.90\%    \\
    radd\_250                                                & 3213                                                     & 4182                                                    & 3942                                                          & 24.77\%    \\
    z4\_268                                                  & 3073                                                     & 3925                                                    & 3619                                                          & 35.92\%    \\
    sym6\_145                                                & 3888                                                     & 4734                                                    & 4632                                                          & 12.06\%    \\
    misex1\_241                                              & 4813                                                     & 5830                                                    & 5734                                                          & 9.44\%     \\
    rd73\_252                                                & 5321                                                     & 7124                                                    & 6446                                                          & 37.60\%    \\
    cycle10\_2\_110                                          & 6050                                                     & 7433                                                    & 7088                                                          & 24.95\%    \\
    square\_root\_7                                          & 7630                                                     & 9250                                                    & 8983                                                          & 16.48\%    \\
    sqn\_258                                                 & 10223                                                    & 13328                                                   & 12176                                                         & 37.10\%    \\
    rd84\_253                                                & 13658                                                    & 17798                                                   & 16856                                                         & 22.75\%    \\
    co14\_215                                                & 17936                                                    & 22814                                                   & 22292                                                         & 10.70\%    \\
    sym9\_193                                                & 34881                                                    & 44544                                                   & 41004                                                         & 36.63\%    \\
    9symml\_195                                              & 34881                                                    & 44544                                                   & 40917                                                         & 37.53\%   \\
    \hline
    \end{tabular}
    \caption{The performance improvement brought by simulated annealing on IBM Q20. Here we compare with the naive mapping $\tau_{nv}: Q \to V$ which maps $q_i$ to $v_i$ for all $i$ in Q20.}
    \label{tab:SA}
\end{table}

The algorithm proposed in \cite{Astar} utilizes $A^*$ to find the best solution of each layer. It has exponential time complexity and only considers one layer for look-ahead when designing the heuristic cost function. Like ours, their $A^*$-based algorithm works for both directed and undirected architecture graphs. As confirmed in \cite{Q20}, it is comparable with the algorithm in \cite{Q20} when Q20 is used as the QPU. So we only make the comparison on QX5. From the experimental results reported in Table~\ref{tab:QX5},  we can see that our algorithm has a conspicuous improvement over the algorithm in \cite{Astar}. Moreover, it is very efficient: for input circuits with up to 10,000 elementary gates, our algorithm finds the solution within one minute.

The algorithm proposed in \cite{Q20} uses reverse traversal technique to search for a good initial mapping and has polynomial complexity. Although it considers multiple levels in its heuristic function, this algorithm does not consider the weights for gates in different layers in the heuristic function.
Unlike our algorithm, the algorithm in \cite{Q20} can only be applied to undirected architecture graphs. Therefore, we only compared it with ours on Q20.
From the experimental results reported in Table~\ref{tab:Q20},  we see that, for small circuits, both algorithms find the optimal output circuits;  but for circuits with large size, our algorithm again has a conspicuous improvement. As for QX5, our algorithm is able to find within two minutes the solution to input circuits with up to 30,000 elementary gates.

We also compared our algorithm with the algorithm proposed in \cite{11}, which also works for both directed and undirected architecture graph and its performance is comparable with the one in \cite{Astar}. In Appendix, from the experimental results reported in Table~5 and 6, we can see that our algorithm also has a better performance.

\begin{table}
    \centering
    \begin{tabular}{cccccc}
    \hline
    Circuit Name     & \begin{tabular}[c]{@{}c@{}}Original\\ Gates\end{tabular} & \begin{tabular}[c]{@{}c@{}}Algorithm \\ in \cite{Astar} \end{tabular} & \begin{tabular}[c]{@{}c@{}}Proposed\\ Algorithm\end{tabular} & \begin{tabular}[c]{@{}c@{}}Running\\ Time(s)\end{tabular} & Improvement \\
    \hline
    mini\_alu\_305     & 173                                                      & 734                                                                & 545                                                          & 0.27                                                      & 33.69\%    \\
    qft\_10            & 200                                                      & 637                                                                & 473                                                          & 0.30                                                      & 37.53\%    \\
    sys6-v0\_111       & 215                                                      & 940                                                                & 701                                                          & 0.32                                                      & 32.97\%    \\
    rd73\_140          & 230                                                      & 934                                                                & 734                                                          & 0.42                                                      & 28.41\%    \\
    sym6\_316          & 270                                                      & 1145                                                               & 925                                                          & 0.43                                                      & 25.14\%    \\
    rd53\_311          & 275                                                      & 1092                                                               & 985                                                          & 0.44                                                      & 13.10\%    \\
    sym9\_146          & 328                                                      & 1317                                                               & 1105                                                         & 0.53                                                      & 21.44\%    \\
    rd84\_142          & 343                                                      & 1381                                                               & 1125                                                         & 0.59                                                      & 24.66\%    \\
    ising\_model\_10   & 480                                                      & 680                                                                & 622                                                          & 0.65                                                      & 29.00\%    \\
    cnt3-5\_180        & 485                                                      & 1703                                                               & 1553                                                         & 0.83                                                      & 12.32\%    \\
    qft\_16            & 512                                                      & 1776                                                               & 1402                                                         & 0.91                                                      & 29.59\%    \\
    ising\_model\_13   & 633                                                      & 913                                                                & 832                                                          & 1.36                                                      & 28.93\%    \\
    ising\_model\_16   & 786                                                      & 1106                                                               & 1049                                                         & 1.64                                                      & 17.81\%    \\
    wim\_266           & 986                                                      & 3867                                                               & 3057                                                         & 1.86                                                      & 28.12\%    \\
    cm152a\_212        & 1221                                                     & 4528                                                               & 3834                                                         & 1.95                                                      & 20.99\%    \\
    cm42a\_207         & 1776                                                     & 6209                                                               & 5612                                                         & 3.39                                                      & 13.47\%    \\
    pm1\_249           & 1776                                                     & 6209                                                               & 5576                                                         & 3.42                                                      & 14.28\%    \\
    dc1\_220           & 1914                                                     & 7009                                                               & 5957                                                         & 3.44                                                      & 20.65\%    \\
    squar5\_261        & 1993                                                     & 7348                                                               & 6641                                                         & 4.60                                                      & 13.20\%    \\
    sqrt8\_260         & 3009                                                     & 11340                                                              & 10181                                                        & 8.05                                                      & 13.91\%    \\
    z4\_268            & 3073                                                     & 11193                                                              & 9995                                                         & 8.40                                                      & 14.75\%    \\
    adr4\_197          & 3439                                                     & 12712                                                              & 11523                                                        & 8.70                                                      & 12.82\%    \\
    sym6\_145          & 3888                                                     & 13426                                                              & 11794                                                        & 10.95                                                     & 17.11\%    \\
    misex1\_241        & 4813                                                     & 17433                                                              & 15714                                                        & 14.74                                                     & 13.62\%    \\
    square\_root\_7    & 7630                                                     & Time Out                                                           & 25972                                                        & 44.05                                                     & \#   \\
    ham15\_107         & 8763                                                     & 31743                                                              & 28829                                                        & 44.66                                                     & 12.68\%    \\
    dc2\_222           & 9462                                                     & 35903                                                              & 32417                                                        & 58.38                                                     & 13.18\%    \\
    sqn\_258           & 10223                                                    & 36957                                                              & 33074                                                        & 58.91                                                     & 14.52\%    \\
    inc\_237           & 10619                                                    & 39151                                                              & 35515                                                        & 59.28                                                     & 12.74\%    \\
    co14\_215          & 17936                                                    & 69830                                                              & 61360                                                        & 320.81                                                    & 16.32\%    \\
    life\_238          & 22445                                                    & 82117                                                              & 75272                                                        & 265.09                                                    & 11.47\%    \\
    max46\_240         & 27126                                                    & 96852                                                              & 88955                                                        & 395.91                                                    & 11.33\%    \\
    9symml\_195        & 34881                                                    & 130153                                                             & 117191                                                       & 667.18                                                    & 13.61\%    \\
    dist\_223          & 38046                                                    & 141729                                                             & 130091                                                       & 829.44                                                    & 11.22\%    \\
    sao2\_257          & 38577                                                    & 146996                                                             & 132110                                                       & 1059.06                                                   & 13.73\%    \\
    plus63mod4096\_163 & 128744                                                   & Time Out                                                           & 445208                                                       & 9669.79                                                   & \#       \\
    Summarizing        & 250896 & 927063 & 836749 & \# & 13.36\% \\
    \hline
    \end{tabular}
    \caption{Comparison of our algorithm with the $A^*$-based algorithm in  \cite{Astar} on IBM QX5.}
    \label{tab:QX5}
\end{table}

\begin{table}
    \centering
    \begin{tabular}{cccccc}
    \hline
    \begin{tabular}[c]{@{}c@{}}Circuit Name\end{tabular} & \begin{tabular}[c]{@{}c@{}}Original\\Gates\end{tabular} &\begin{tabular}[c]{@{}c@{}}Algorithm \\ in \cite{Q20}\end{tabular}  &\begin{tabular}[c]{@{}c@{}}Proposed\\Algorithm\end{tabular} &\begin{tabular}[c]{@{}c@{}}Running\\Time(s)\end{tabular} & Improvement \\
    \hline
    4mod5-v1\_22                                             & 21             & 21                       & 21                 & 0.00               & 0.00\%      \\
    mod5mils\_65                                             & 35             & 35                       & 35                 & 0.00     & 0.00\%      \\
    alu-v0\_27                                               & 36             & 39                       & 42                 & 0.00     & -75.00\%    \\
    decod24-v2 43                                            & 52             & 52                       & 52                 & 0.00    & 0.00\%      \\
    4gt13\_92                                                & 66             & 66                       & 66                 & 0.00     & 0.00\%      \\
    ising\_model\_10                                         & 480            & 480                      & 480                & 0.00    & 0.00\%      \\
    ising\_model\_13                                         & 633            & 633                      & 633                & 0.01     & 0.00\%      \\
    ising\_model\_16                                         & 786            & 786                      & 786                & 0.01    & 0.00\%      \\
    qft\_10                                                  & 200            & 254                      & 236                & 0.16     & 32.73\%     \\
    qft\_16                                                  & 512            & 698                      & 647                & 0.38     & 27.27\%     \\
    rd84\_142                                                & 343            & 448                      & 445                & 1.50     & 2.83\%      \\
    adr4\_197                                                & 3439           & 5053                     & 4150               & 2.08     & 55.91\%     \\
    radd\_250                                                & 3213           & 4488                     & 3942               & 2.13     & 42.79\%     \\
    z4\_268                                                  & 3073           & 4438                     & 3619               & 2.65     & 59.96\%     \\
    sym6\_145                                                & 3888           & 5160                     & 4632               & 2.82      & 41.48\%     \\
    misex1\_241                                              & 4813           & 6334                     & 5734               & 3.44     & 39.42\%     \\
    rd73\_252                                                & 5321           & 7454                     & 6446               & 5.21     & 47.24\%     \\
    cycle10\_2\_110                                          & 6050           & 8672                     & 7088               & 5.46     & 60.39\%     \\
    square\_root\_7                                          & 7630           & 10228                    & 8983               & 12.24     & 47.90\%     \\
    sqn\_258                                                 & 10223          & 14567                    & 12176              & 13.91     & 55.03\%     \\
    rd84\_253                                                & 13658          & 19805                    & 16856              & 34.75     & 47.97\%     \\
    co14\_215                                                & 17936          & 26918                    & 22292              & 88.90      & 51.50\%     \\
    sym9\_193                                                & 34881          & 51534                    & 41004              & 126.68     & 63.23\%     \\
    9symml\_195                                              & 34881          & 52149                    & 40917              & 137.47     & 65.04\%    \\
    Summarizing & 152170 & 220312 & 181282 & \# & 57.28\% \\
    \hline
    \end{tabular}
    \caption{Comparison of our algorithm with the algorithm in  \cite{Q20} on IBM Q20.}
    \label{tab:Q20}
\end{table}

\begin{figure}
    \centering
    \includegraphics[width=0.6\textwidth]{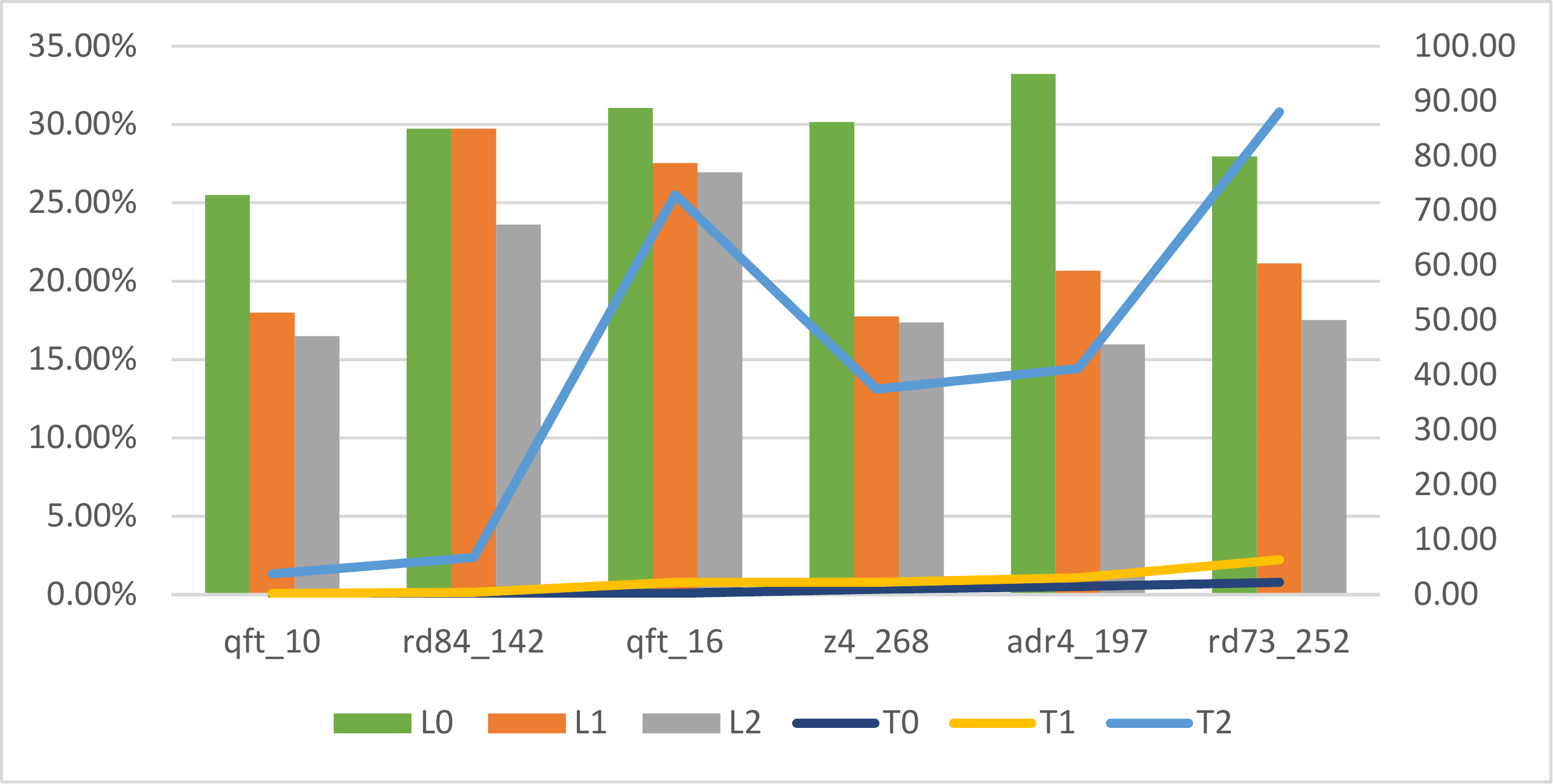}
    \caption{Experiments on IBM Q20 for different look-ahead depths. The vertical axis on the left is the ratio of the number of
    added gates to that of the original gates, and the right axis shows the consumed time. The bar and line represent respectively the number of gates and the consumed time for different circuits and different look-ahead depths.}
    \label{fig:depth}
\end{figure}

It is worth mentioning that, if the depth for look-ahead in the selection process is increased, the quality of the output circuits could be further improved. However, the time consumption will be increased dramatically. See Fig.~\ref{fig:depth} for the experiment on a few examples, which indicates that 1-depth look-ahead reaches the best trade off of time and performance. Although the time overhead for increasing the depth is vast, it may still be acceptable in some application scenarios and this can easily be done by adjusting the relevant parameters of our algorithm. Besides, the weight parameters in the heuristic function are also adjustable when different architecture graphs and circuits are considered.

\section{Conclusion}
\label{sec:conclusion}


In this paper, we propose an algorithm to solve the quantum circuit transformation problem by using simulated annealing and heuristic search. A double look-ahead mechanism is novelly adopted in the algorithm. We look ahead at subsequent layers when defining a flexible heuristic cost function which also supports weight parameters to reflect the variable influence of gates in different layers. Moreover, we look ahead at grandchild states with minimal cost in selecting the best state for the next step of the circuit transformation.
Detailed evaluation on extensive realistic circuits shows that our algorithm has consistent and significant  improvement when compared with the two state-of-the-art algorithms proposed in the literature for IBM QX5 and Q20.

For future studies, we propose the following problems to solve. First, our program still runs slowly for circuits with large sizes. Thus it is necessary to optimize the code to reduce the running time. Second, the quality of the initial mappings obtained from the simulated annealing algorithm (Alg.~\ref{alg:initial map}) is not stable, which is not acceptable for commercial use. Third, we only consider connectivity in the architecture graphs; other constraints like cross talk, gate error and qubits decoherence should be included to make the algorithm more practical. Fourth, only using the sizes of circuits as the criterion for evaluation is not enough. Criteria like circuit error and running time should also be considered in future work.


\newpage
\appendix
\section{More experimental results}
\begin{longtable}{cccccc}
\hline
Circuit Name& \begin{tabular}[c]{@{}c@{}}Original\\ Gates\end{tabular} & \begin{tabular}[c]{@{}c@{}}Algorithm\\ in \cite{11}\end{tabular} & \begin{tabular}[c]{@{}c@{}}Proposed\\ Algorithm\end{tabular} & \begin{tabular}[c]{@{}c@{}}Running\\ Time(s)\end{tabular} & Improvement \\
\hline
\endhead
graycode6\_47         & 5              & 13                       & 13                 & 0.00            & 0.00\%     \\
xor5\_254             & 7              & 30                       & 14                 & 0.00            & 69.57\%    \\
ex1\_226              & 7              & 30                       & 21                 & 0.00            & 39.13\%    \\
4gt11\_84             & 18             & 55                       & 47                 & 0.01            & 21.62\%    \\
ex-1\_166             & 19             & 68                       & 48                 & 0.01            & 40.82\%    \\
ham3\_102             & 20             & 76                       & 52                 & 0.01            & 42.86\%    \\
4mod5-v0\_20          & 20             & 61                       & 49                 & 0.01            & 29.27\%    \\
4mod5-v1\_22          & 21             & 62                       & 53                 & 0.01            & 21.95\%    \\
mod5d1\_63            & 22             & 78                       & 58                 & 0.02            & 35.71\%    \\
4gt11\_83             & 23             & 82                       & 63                 & 0.02            & 32.20\%    \\
4gt11\_82             & 27             & 109                      & 78                 & 0.03            & 37.80\%    \\
rd32-v0\_66           & 34             & 108                      & 91                 & 0.02            & 22.97\%    \\
mod5mils\_65          & 35             & 132                      & 88                 & 0.02            & 45.36\%    \\
4mod5-v0\_19          & 35             & 128                      & 95                 & 0.02            & 35.48\%    \\
rd32-v1\_68           & 36             & 110                      & 93                 & 0.02            & 22.97\%    \\
alu-v0\_27            & 36             & 118                      & 97                 & 0.02            & 25.61\%    \\
3\_17\_13             & 36             & 102                      & 93                 & 0.02            & 13.64\%    \\
4mod5-v1\_24          & 36             & 113                      & 96                 & 0.02            & 22.08\%    \\
alu-v1\_29            & 37             & 119                      & 98                 & 0.02            & 25.61\%    \\
alu-v1\_28            & 37             & 131                      & 105                & 0.02            & 27.66\%    \\
alu-v3\_35            & 37             & 119                      & 106                & 0.02            & 15.85\%    \\
alu-v2\_33            & 37             & 119                      & 91                 & 0.02            & 34.15\%    \\
alu-v4\_37            & 37             & 119                      & 98                 & 0.02            & 25.61\%    \\
miller\_11            & 50             & 169                      & 139                & 0.03            & 25.21\%    \\
decod24-v0\_38        & 51             & 154                      & 126                & 0.03            & 27.18\%    \\
alu-v3\_34            & 52             & 181                      & 160                & 0.03            & 16.28\%    \\
decod24-v2\_43        & 52             & 155                      & 141                & 0.02            & 13.59\%    \\
mod5d2\_64            & 53             & 162                      & 163                & 0.03            & -0.92\%    \\
4gt13\_92             & 66             & 218                      & 187                & 0.04            & 20.39\%    \\
4gt13-v1\_93          & 68             & 212                      & 186                & 0.05            & 18.06\%    \\
one-two-three-v2\_100 & 69             & 217                      & 199                & 0.04            & 12.16\%    \\
4mod5-v1\_23          & 69             & 234                      & 214                & 0.05            & 12.12\%    \\
4mod5-v0\_18          & 69             & 215                      & 212                & 0.05            & 2.05\%     \\
one-two-three-v3\_101 & 70             & 244                      & 199                & 0.06            & 25.86\%    \\
4mod5-bdd\_287        & 70             & 219                      & 200                & 0.04            & 12.75\%    \\
decod24-bdd\_294      & 73             & 216                      & 206                & 0.04            & 6.99\%     \\
4gt5\_75              & 83             & 263                      & 226                & 0.05            & 20.56\%    \\
alu-v0\_26            & 84             & 281                      & 248                & 0.06            & 16.75\%    \\
rd32\_270             & 84             & 264                      & 245                & 0.05            & 10.56\%    \\
alu-bdd\_288          & 84             & 245                      & 247                & 0.07            & -1.24\%    \\
decod24-v1\_41        & 85             & 281                      & 245                & 0.05            & 18.37\%    \\
4gt5\_76              & 91             & 302                      & 271                & 0.08            & 14.69\%    \\
4gt13\_91             & 103            & 348                      & 293                & 0.08            & 22.45\%    \\
4gt13\_90             & 107            & 371                      & 319                & 0.09            & 19.70\%    \\
alu-v4\_36            & 115            & 375                      & 326                & 0.08            & 18.85\%    \\
4gt5\_77              & 131            & 409                      & 363                & 0.10            & 16.55\%    \\
one-two-three-v1\_99  & 132            & 447                      & 392                & 0.09            & 17.46\%    \\
rd53\_138             & 132            & 412                      & 381                & 0.22            & 11.07\%    \\
one-two-three-v0\_98  & 146            & 445                      & 428                & 0.12            & 5.69\%     \\
4gt10-v1\_81          & 148            & 506                      & 422                & 0.10            & 23.46\%    \\
decod24-v3\_45        & 150            & 496                      & 407                & 0.11            & 25.72\%    \\
aj-e11\_165           & 151            & 472                      & 409                & 0.11            & 19.63\%    \\
4mod7-v0\_94          & 162            & 540                      & 456                & 0.14            & 22.22\%    \\
alu-v2\_32            & 163            & 517                      & 472                & 0.13            & 12.71\%    \\
4mod7-v1\_96          & 164            & 496                      & 485                & 0.12            & 3.31\%     \\
cnt3-5\_179           & 175            & 842                      & 517                & 0.59            & 48.73\%    \\
mod10\_176            & 178            & 566                      & 498                & 0.15            & 17.53\%    \\
4gt4-v0\_80           & 179            & 596                      & 504                & 0.11            & 22.06\%    \\
4gt12-v0\_88          & 194            & 644                      & 567                & 0.14            & 17.11\%    \\
0410184\_169          & 211            & 877                      & 664                & 0.61            & 31.98\%    \\
4\_49\_16             & 217            & 725                      & 612                & 0.16            & 22.24\%    \\
4gt12-v1\_89          & 228            & 751                      & 669                & 0.19            & 15.68\%    \\
4gt4-v0\_79           & 231            & 720                      & 668                & 0.22            & 10.63\%    \\
hwb4\_49              & 233            & 745                      & 684                & 0.20            & 11.91\%    \\
4gt4-v0\_78           & 235            & 739                      & 683                & 0.19            & 11.11\%    \\
mod10\_171            & 244            & 779                      & 696                & 0.23            & 15.51\%    \\
4gt12-v0\_87          & 247            & 786                      & 714                & 0.21            & 13.36\%    \\
4gt12-v0\_86          & 251            & 805                      & 729                & 0.26            & 13.72\%    \\
4gt4-v0\_72           & 258            & 818                      & 741                & 0.27            & 13.75\%    \\
4gt4-v1\_74           & 273            & 900                      & 801                & 0.22            & 15.79\%    \\
mini-alu\_167         & 288            & 919                      & 836                & 0.24            & 13.15\%    \\
one-two-three-v0\_97  & 290            & 861                      & 853                & 0.23            & 1.40\%     \\
rd53\_135             & 296            & 1029                     & 919                & 0.36            & 15.01\%    \\
ham7\_104             & 320            & 1123                     & 979                & 0.30            & 17.93\%    \\
decod24-enable\_126   & 338            & 1091                     & 1005               & 0.30            & 11.42\%    \\
mod8-10\_178          & 342            & 1228                     & 1027               & 0.32            & 22.69\%    \\
4gt4-v0\_73           & 395            & 1289                     & 1153               & 0.43            & 15.21\%    \\
ex3\_229              & 403            & 1247                     & 1166               & 0.37            & 9.60\%     \\
mod8-10\_177          & 440            & 1401                     & 1299               & 0.46            & 10.61\%    \\
alu-v2\_31            & 451            & 1458                     & 1296               & 0.42            & 16.09\%    \\
C17\_204              & 467            & 1588                     & 1439               & 0.54            & 13.29\%    \\
rd53\_131             & 469            & 1614                     & 1394               & 0.52            & 19.21\%    \\
alu-v2\_30            & 504            & 1627                     & 1515               & 0.50            & 9.97\%     \\
mod5adder\_127        & 555            & 1758                     & 1576               & 0.60            & 15.13\%    \\
rd53\_133             & 580            & 1954                     & 1743               & 0.84            & 15.36\%    \\
majority\_239         & 612            & 2073                     & 1791               & 0.64            & 19.30\%    \\
ex2\_227              & 631            & 2130                     & 1895               & 0.73            & 15.68\%    \\
cm82a\_208            & 650            & 2093                     & 2020               & 0.93            & 5.06\%     \\
sf\_276               & 778            & 2481                     & 2241               & 0.74            & 14.09\%    \\
sf\_274               & 781            & 2508                     & 2275               & 0.87            & 13.49\%    \\
con1\_216             & 954            & 3232                     & 3040               & 1.50            & 8.43\%     \\
rd53\_130             & 1043           & 3418                     & 3110               & 1.39            & 12.97\%    \\
f2\_232               & 1206           & 3887                     & 3627               & 1.65            & 9.70\%     \\
rd53\_251             & 1291           & 4435                     & 3867               & 1.77            & 18.07\%    \\
hwb5\_53              & 1336           & 4462                     & 3942               & 1.52            & 16.63\%    \\
radd\_250             & 3213           & 11330                    & 10440              & 8.81            & 10.96\%    \\
rd73\_252             & 5321           & 18670                    & 17335              & 18.87           & 10.00\%    \\
cycle10\_2\_110       & 6050           & 21704                    & 19699              & 25.44           & 12.81\%    \\
hwb6\_56              & 6723           & 22502                    & 20571              & 22.13           & 12.24\%    \\
cm85a\_209            & 11414          & 41785                    & 38616              & 72.29           & 10.43\%    \\
rd84\_253             & 13658          & 50103                    & 46284              & 115.99          & 10.48\%    \\
root\_255             & 17159          & 61424                    & 58297              & 204.43          & 7.06\%     \\
mlp4\_245             & 18852          & 70980                    & 63717              & 205.33          & 13.93\%    \\
urf2\_277             & 20112          & 78710                    & 70143              & 241.43          & 14.62\%    \\
sym9\_148             & 21504          & 73234                    & 67407              & 211.39          & 11.26\%    \\
hwb7\_59              & 24379          & 82058                    & 75605              & 240.25          & 11.19\%    \\
clip\_206             & 33827          & 125443                   & 115217             & 663.92          & 11.16\%    \\
sym9\_193             & 34881          & 125917                   & 117105             & 651.67          & 9.68\%     \\
dist\_223             & 38046          & 137543                   & 130271             & 829.44          & 7.31\%     \\
sao2\_257             & 38577          & 145946                   & 132533             & 989.69          & 12.49\%    \\
urf5\_280             & 49829          & 183656                   & 172758             & 1216.33         & 8.14\%     \\
urf1\_278             & 54766          & 208475                   & 196052             & 1666.36         & 8.08\%     \\
sym10\_262            & 64283          & 235802                   & 218264             & 2467.67         & 10.23\%    \\
Summarizing & 485117 & 1769629 & 1636683 & \# & 10.35\% \\
\hline
\caption{Comparison between our algorithm and the algorithm in \cite{11} on IBM QX5, where ratio in the Comparison column is obtained by (added gates in our program) vs. (added gates in their program)}
\end{longtable}

\begin{longtable}{cccccc}
\hline
Circuit Name          & \begin{tabular}[c]{@{}c@{}}Original\\ Gates\end{tabular} & \begin{tabular}[c]{@{}c@{}}Algorithm\\ in \cite{11}\end{tabular} & \begin{tabular}[c]{@{}c@{}}Proposed\\ Algorithm\end{tabular} & \begin{tabular}[c]{@{}c@{}}Running\\ Time(s)\end{tabular} & Improvement \\
\hline
\endhead
graycode6\_47         & 5                                                        & 5                                                                  & 5                                                            & 0.00                                                      & 0.00\%      \\
xor5\_254             & 7                                                        & 7                                                                  & 7                                                            & 0.00                                                      & 0.00\%      \\
ex1\_226              & 7                                                        & 7                                                                  & 7                                                            & 0.00                                                      & 0.00\%      \\
4gt11\_84             & 18                                                       & 18                                                                 & 18                                                           & 0.00                                                      & 0.00\%      \\
ex-1\_166             & 19                                                       & 19                                                                 & 19                                                           & 0.00                                                      & 0.00\%      \\
ham3\_102             & 20                                                       & 20                                                                 & 20                                                           & 0.00                                                      & 0.00\%      \\
4mod5-v0\_20          & 20                                                       & 29                                                                 & 20                                                           & 0.00                                                      & 90.00\%     \\
4mod5-v1\_22          & 21                                                       & 30                                                                 & 21                                                           & 0.00                                                      & 90.00\%     \\
mod5d1\_63            & 22                                                       & 22                                                                 & 22                                                           & 0.00                                                      & 0.00\%      \\
4gt11\_83             & 23                                                       & 35                                                                 & 23                                                           & 0.00                                                      & 92.31\%     \\
4gt11\_82             & 27                                                       & 39                                                                 & 30                                                           & 0.00                                                      & 69.23\%     \\
rd32-v0\_66           & 34                                                       & 34                                                                 & 34                                                           & 0.00                                                      & 0.00\%      \\
mod5mils\_65          & 35                                                       & 44                                                                 & 35                                                           & 0.00                                                      & 90.00\%     \\
4mod5-v0\_19          & 35                                                       & 44                                                                 & 35                                                           & 0.00                                                      & 90.00\%     \\
rd32-v1\_68           & 36                                                       & 36                                                                 & 36                                                           & 0.00                                                      & 0.00\%      \\
alu-v0\_27            & 36                                                       & 39                                                                 & 42                                                           & 0.01                                                      & -75.00\%    \\
3\_17\_13             & 36                                                       & 36                                                                 & 36                                                           & 0.00                                                      & 0.00\%      \\
4mod5-v1\_24          & 36                                                       & 48                                                                 & 36                                                           & 0.00                                                      & 92.31\%     \\
alu-v1\_29            & 37                                                       & 40                                                                 & 43                                                           & 0.01                                                      & -75.00\%    \\
alu-v1\_28            & 37                                                       & 40                                                                 & 43                                                           & 0.01                                                      & -75.00\%    \\
alu-v3\_35            & 37                                                       & 40                                                                 & 43                                                           & 0.01                                                      & -75.00\%    \\
alu-v2\_33            & 37                                                       & 46                                                                 & 43                                                           & 0.01                                                      & 30.00\%     \\
alu-v4\_37            & 37                                                       & 40                                                                 & 43                                                           & 0.01                                                      & -75.00\%    \\
miller\_11            & 50                                                       & 50                                                                 & 50                                                           & 0.00                                                      & 0.00\%      \\
decod24-v0\_38        & 51                                                       & 51                                                                 & 51                                                           & 0.00                                                      & 0.00\%      \\
alu-v3\_34            & 52                                                       & 55                                                                 & 58                                                           & 0.01                                                      & -75.00\%    \\
decod24-v2\_43        & 52                                                       & 52                                                                 & 52                                                           & 0.00                                                      & 0.00\%      \\
mod5d2\_64            & 53                                                       & 65                                                                 & 65                                                           & 0.01                                                      & 0.00\%      \\
4gt13\_92             & 66                                                       & 84                                                                 & 66                                                           & 0.00                                                      & 94.74\%     \\
4gt13-v1\_93          & 68                                                       & 86                                                                 & 68                                                           & 0.00                                                      & 94.74\%     \\
one-two-three-v2\_100 & 69                                                       & 78                                                                 & 78                                                           & 0.01                                                      & 0.00\%      \\
4mod5-v1\_23          & 69                                                       & 81                                                                 & 78                                                           & 0.02                                                      & 23.08\%     \\
4mod5-v0\_18          & 69                                                       & 78                                                                 & 78                                                           & 0.01                                                      & 0.00\%      \\
one-two-three-v3\_101 & 70                                                       & 85                                                                 & 76                                                           & 0.01                                                      & 56.25\%     \\
4mod5-bdd\_287        & 70                                                       & 85                                                                 & 76                                                           & 0.01                                                      & 56.25\%     \\
decod24-bdd\_294      & 73                                                       & 94                                                                 & 88                                                           & 0.02                                                      & 27.27\%     \\
4gt5\_75              & 83                                                       & 98                                                                 & 98                                                           & 0.02                                                      & 0.00\%      \\
alu-v0\_26            & 84                                                       & 105                                                                & 93                                                           & 0.01                                                      & 54.55\%     \\
rd32\_270             & 84                                                       & 102                                                                & 96                                                           & 0.01                                                      & 31.58\%     \\
alu-bdd\_288          & 84                                                       & 129                                                                & 108                                                          & 0.03                                                      & 45.65\%     \\
decod24-v1\_41        & 85                                                       & 103                                                                & 100                                                          & 0.02                                                      & 15.79\%     \\
4gt5\_76              & 91                                                       & 118                                                                & 106                                                          & 0.02                                                      & 42.86\%     \\
4gt13\_91             & 103                                                      & 109                                                                & 118                                                          & 0.02                                                      & -128.57\%   \\
4gt13\_90             & 107                                                      & 116                                                                & 134                                                          & 0.02                                                      & -180.00\%   \\
alu-v4\_36            & 115                                                      & 151                                                                & 130                                                          & 0.02                                                      & 56.76\%     \\
4gt5\_77              & 131                                                      & 167                                                                & 140                                                          & 0.04                                                      & 72.97\%     \\
one-two-three-v1\_99  & 132                                                      & 171                                                                & 144                                                          & 0.02                                                      & 67.50\%     \\
rd53\_138             & 132                                                      & 171                                                                & 159                                                          & 0.05                                                      & 30.00\%     \\
one-two-three-v0\_98  & 146                                                      & 173                                                                & 170                                                          & 0.04                                                      & 10.71\%     \\
4gt10-v1\_81          & 148                                                      & 181                                                                & 175                                                          & 0.04                                                      & 17.65\%     \\
decod24-v3\_45        & 150                                                      & 189                                                                & 165                                                          & 0.04                                                      & 60.00\%     \\
aj-e11\_165           & 151                                                      & 175                                                                & 169                                                          & 0.02                                                      & 24.00\%     \\
4mod7-v0\_94          & 162                                                      & 201                                                                & 174                                                          & 0.04                                                      & 67.50\%     \\
alu-v2\_32            & 163                                                      & 202                                                                & 178                                                          & 0.02                                                      & 60.00\%     \\
4mod7-v1\_96          & 164                                                      & 206                                                                & 182                                                          & 0.03                                                      & 55.81\%     \\
cnt3-5\_179           & 175                                                      & 262                                                                & 190                                                          & 0.04                                                      & 81.82\%     \\
mod10\_176            & 178                                                      & 214                                                                & 202                                                          & 0.03                                                      & 32.43\%     \\
4gt4-v0\_80           & 179                                                      & 257                                                                & 203                                                          & 0.05                                                      & 68.35\%     \\
4gt12-v0\_88          & 194                                                      & 215                                                                & 215                                                          & 0.05                                                      & 0.00\%      \\
0410184\_169          & 211                                                      & 286                                                                & 223                                                          & 0.02                                                      & 82.89\%     \\
4\_49\_16             & 217                                                      & 286                                                                & 253                                                          & 0.04                                                      & 47.14\%     \\
4gt12-v1\_89          & 228                                                      & 321                                                                & 252                                                          & 0.03                                                      & 73.40\%     \\
4gt4-v0\_79           & 231                                                      & 327                                                                & 243                                                          & 0.04                                                      & 86.60\%     \\
hwb4\_49              & 233                                                      & 278                                                                & 266                                                          & 0.04                                                      & 26.09\%     \\
4gt4-v0\_78           & 235                                                      & 334                                                                & 250                                                          & 0.05                                                      & 84.00\%     \\
mod10\_171            & 244                                                      & 304                                                                & 268                                                          & 0.04                                                      & 59.02\%     \\
4gt12-v0\_87          & 247                                                      & 370                                                                & 253                                                          & 0.01                                                      & 94.35\%     \\
4gt12-v0\_86          & 251                                                      & 374                                                                & 260                                                          & 0.02                                                      & 91.94\%     \\
4gt4-v0\_72           & 258                                                      & 348                                                                & 300                                                          & 0.05                                                      & 52.75\%     \\
4gt4-v1\_74           & 273                                                      & 387                                                                & 351                                                          & 0.10                                                      & 31.30\%     \\
mini-alu\_167         & 288                                                      & 363                                                                & 321                                                          & 0.06                                                      & 55.26\%     \\
one-two-three-v0\_97  & 290                                                      & 356                                                                & 356                                                          & 0.09                                                      & 0.00\%      \\
rd53\_135             & 296                                                      & 344                                                                & 350                                                          & 0.11                                                      & -12.24\%    \\
ham7\_104             & 320                                                      & 422                                                                & 401                                                          & 0.08                                                      & 20.39\%     \\
decod24-enable\_126   & 338                                                      & 419                                                                & 425                                                          & 0.14                                                      & -7.32\%     \\
mod8-10\_178          & 342                                                      & 504                                                                & 363                                                          & 0.04                                                      & 86.50\%     \\
4gt4-v0\_73           & 395                                                      & 572                                                                & 437                                                          & 0.05                                                      & 75.84\%     \\
ex3\_229              & 403                                                      & 577                                                                & 421                                                          & 0.05                                                      & 89.14\%     \\
mod8-10\_177          & 440                                                      & 575                                                                & 479                                                          & 0.05                                                      & 70.59\%     \\
alu-v2\_31            & 451                                                      & 514                                                                & 505                                                          & 0.07                                                      & 14.06\%     \\
C17\_204              & 467                                                      & 581                                                                & 563                                                          & 0.14                                                      & 15.65\%     \\
rd53\_131             & 469                                                      & 556                                                                & 559                                                          & 0.17                                                      & -3.41\%     \\
alu-v2\_30            & 504                                                      & 609                                                                & 549                                                          & 0.07                                                      & 56.60\%     \\
mod5adder\_127        & 555                                                      & 642                                                                & 606                                                          & 0.11                                                      & 40.91\%     \\
rd53\_133             & 580                                                      & 739                                                                & 685                                                          & 0.16                                                      & 33.75\%     \\
majority\_239         & 612                                                      & 735                                                                & 696                                                          & 0.13                                                      & 31.45\%     \\
ex2\_227              & 631                                                      & 901                                                                & 727                                                          & 0.22                                                      & 64.21\%     \\
cm82a\_208            & 650                                                      & 872                                                                & 734                                                          & 0.14                                                      & 61.88\%     \\
sf\_276               & 778                                                      & 1162                                                               & 802                                                          & 0.05                                                      & 93.51\%     \\
sf\_274               & 781                                                      & 1162                                                               & 805                                                          & 0.04                                                      & 93.46\%     \\
con1\_216             & 954                                                      & 1329                                                               & 1146                                                         & 0.39                                                      & 48.67\%     \\
rd53\_130             & 1043                                                     & 1433                                                               & 1214                                                         & 0.31                                                      & 56.01\%     \\
f2\_232               & 1206                                                     & 1431                                                               & 1419                                                         & 0.43                                                      & 5.31\%      \\
rd53\_251             & 1291                                                     & 1600                                                               & 1495                                                         & 0.34                                                      & 33.87\%     \\
hwb5\_53              & 1336                                                     & 1546                                                               & 1510                                                         & 0.30                                                      & 17.06\%     \\
radd\_250             & 3213                                                     & 4860                                                               & 3882                                                         & 2.34                                                      & 59.34\%     \\
rd73\_252             & 5321                                                     & 7436                                                               & 6386                                                         & 5.31                                                      & 49.62\%     \\
cycle10\_2\_110       & 6050                                                     & 8474                                                               & 7346                                                         & 6.25                                                      & 46.52\%     \\
hwb6\_56              & 6723                                                     & 8442                                                               & 7827                                                         & 3.88                                                      & 35.76\%     \\
cm85a\_209            & 11414                                                    & 15587                                                              & 13751                                                        & 16.67                                                     & 43.99\%     \\
rd84\_253             & 13658                                                    & 18944                                                              & 16904                                                        & 37.18                                                     & 38.59\%     \\
root\_255             & 17159                                                    & 22760                                                              & 20684                                                        & 48.44                                                     & 37.06\%     \\
mlp4\_245             & 18852                                                    & 25314                                                              & 22968                                                        & 66.49                                                     & 36.30\%     \\
urf2\_277             & 20112                                                    & 28317                                                              & 26046                                                        & 77.04                                                     & 27.67\%     \\
sym9\_148             & 21504                                                    & 27942                                                              & 23676                                                        & 31.55                                                     & 66.25\%     \\
hwb7\_59              & 24379                                                    & 30757                                                              & 28981                                                        & 63.92                                                     & 27.84\%     \\
clip\_206             & 33827                                                    & 46451                                                              & 40670                                                        & 173.59                                                    & 45.79\%     \\
sym9\_193             & 34881                                                    & 46335                                                              & 41322                                                        & 144.55                                                    & 43.76\%     \\
dist\_223             & 38046                                                    & 50880                                                              & 44982                                                        & 187.64                                                    & 45.95\%     \\
sao2\_257             & 38577                                                    & 50319                                                              & 46404                                                        & 233.36                                                    & 33.34\%     \\
urf5\_280             & 49829                                                    & 70265                                                              & 62894                                                        & 374.76                                                    & 36.07\%     \\
urf1\_278             & 54766                                                    & 79366                                                              & 70444                                                        & 649.03                                                    & 36.27\%     \\
sym10\_262            & 64283                                                    & 84398                                                              & 75980                                                        & 518.75                                                    & 41.85\%     \\
hwb8\_113             & 24379                                                    & 104756                                                             & 84356                                                        & 646.41                                                    & 25.38\%     \\ 
Summarizing & 509496 & 760639 & 670984 & \# & 35.70\%\\
\hline
\caption{Comparison between our algorithm and the algorithm in \cite{11} on IBM Q20, where ratio in the Comparison column is obtained by (added gates in our program) vs. (added gates in their program)}
\label{tab:my-table}\\
\end{longtable}

\bibliographystyle{plain}
\bibliography{references}

\begin{thebibliography}{10}

\bibitem{qiskit}
Gadi Aleksandrowicz, Thomas Alexander, P~Barkoutsos, L~Bello, Y~Ben-Haim,
  D~Bucher, FJ~Cabrera-Hern{\'a}ndez, J~Carballo-Franquis, A~Chen, CF~Chen,
  et~al.
\newblock Qiskit: An open-source framework for quantum computing.
\newblock {\em Accessed on: Mar}, 16, 2019.

\bibitem{matching}
Noga Alon, Fan~RK Chung, and Ronald~L Graham.
\newblock Routing permutations on graphs via matchings.
\newblock {\em SIAM journal on discrete mathematics}, 7(3):513--530, 1994.

\bibitem{10}
Adriano Barenco, Charles~H Bennett, Richard Cleve, David~P DiVincenzo, Norman
  Margolus, Peter Shor, Tycho Sleator, John~A Smolin, and Harald Weinfurter.
\newblock Elementary gates for quantum computation.
\newblock {\em Physical review A}, 52(5):3457, 1995.

\bibitem{planning}
Kyle~EC Booth, Minh Do, J~Christopher Beck, Eleanor Rieffel, Davide Venturelli,
  and Jeremy Frank.
\newblock Comparing and integrating constraint programming and temporal
  planning for quantum circuit compilation.
\newblock In {\em Twenty-Eighth International Conference on Automated Planning
  and Scheduling}, 2018.

\bibitem{complexity}
Adi Botea, Akihiro Kishimoto, and Radu Marinescu.
\newblock On the complexity of quantum circuit compilation.
\newblock In {\em Eleventh Annual Symposium on Combinatorial Search}, 2018.

\bibitem{6}
Andrew~M Childs, Eddie Schoute, and Cem~M Unsal.
\newblock Circuit transformations for quantum architectures.
\newblock {\em arXiv preprint arXiv:1902.09102}, 2019.

\bibitem{11}
Alexander Cowtan, Silas Dilkes, Ross Duncan, Alexandre Krajenbrink, Will
  Simmons, and Seyon Sivarajah.
\newblock On the qubit routing problem.
\newblock {\em arXiv preprint arXiv:1902.08091}, 2019.

\bibitem{13}
Will Finigan, Michael Cubeddu, Thomas Lively, Johannes Flick, and Prineha
  Narang.
\newblock Qubit allocation for noisy intermediate-scale quantum computers.
\newblock {\em arXiv preprint arXiv:1810.08291}, 2018.

\bibitem{compile}
Thomas H{\"a}ner, Damian~S Steiger, Krysta Svore, and Matthias Troyer.
\newblock A software methodology for compiling quantum programs.
\newblock {\em Quantum Science and Technology}, 3(2):020501, 2018.

\bibitem{DG}
Toshinari Itoko, Rudy Raymond, Takashi Imamichi, Atsushi Matsuo, and Andrew~W
  Cross.
\newblock Quantum circuit compilers using gate commutation rules.
\newblock In {\em Proceedings of the 24th Asia and South Pacific Design
  Automation Conference}, pages 191--196. ACM, 2019.

\bibitem{annealing}
Scott Kirkpatrick, C~Daniel Gelatt, and Mario~P Vecchi.
\newblock Optimization by simulated annealing.
\newblock {\em science}, 220(4598):671--680, 1983.

\bibitem{Ud2}
Aleks Kissinger and Arianne Meijer-van de~Griend.
\newblock Cnot circuit extraction for topologically-constrained quantum
  memories.
\newblock {\em arXiv preprint arXiv:1904.00633}, 2019.

\bibitem{Q20}
Gushu Li, Yufei Ding, and Yuan Xie.
\newblock Tackling the qubit mapping problem for nisq-era quantum devices.
\newblock In {\em Proceedings of the Twenty-Fourth International Conference on
  Architectural Support for Programming Languages and Operating Systems}, pages
  1001--1014. ACM, 2019.

\bibitem{token}
Tillmann Miltzow, Lothar Narins, Yoshio Okamoto, G{\"u}nter Rote, Antonis
  Thomas, and Takeaki Uno.
\newblock Approximation and hardness for token swapping.
\newblock {\em arXiv preprint arXiv:1602.05150}, 2016.

\bibitem{qubitmovement}
Prakash Murali, Jonathan~M Baker, Ali Javadi-Abhari, Frederic~T Chong, and
  Margaret Martonosi.
\newblock Noise-adaptive compiler mappings for noisy intermediate-scale quantum
  computers.
\newblock In {\em Proceedings of the Twenty-Fourth International Conference on
  Architectural Support for Programming Languages and Operating Systems}, pages
  1015--1029. ACM, 2019.

\bibitem{Ud1}
Beatrice Nash, Vlad Gheorghiu, and Michele Mosca.
\newblock Quantum circuit optimizations for nisq architectures.
\newblock {\em arXiv preprint arXiv:1904.01972}, 2019.

\bibitem{1}
Michael~A Nielsen and Isaac~L Chuang.
\newblock Quantum information and quantum computation.
\newblock {\em Cambridge: Cambridge University Press}, 2(8):23, 2000.

\bibitem{Rodney}
Shin Nishio, Yulu Pan, Takahiko Satoh, Hideharu Amano, and Rodney Van~Meter.
\newblock Extracting success from ibm's 20-qubit machines using error-aware
  compilation.
\newblock {\em arXiv preprint arXiv:1903.10963}, 2019.

\bibitem{initialmap}
Alexandru Paler.
\newblock On the influence of initial qubit placement during nisq circuit
  compilation.
\newblock In {\em International Workshop on Quantum Technology and Optimization
  Problems}, pages 207--217. Springer, 2019.

\bibitem{NISQ}
John Preskill.
\newblock Quantum computing in the nisq era and beyond.
\newblock {\em Quantum}, 2:79, 2018.

\bibitem{7}
Marcos~Yukio Siraichi, Vin{\'\i}cius Fernandes~dos Santos, Sylvain Collange,
  and Fernando Magno~Quint{\~a}o Pereira.
\newblock Qubit allocation.
\newblock In {\em Proceedings of the 2018 International Symposium on Code
  Generation and Optimization}, pages 113--125. ACM, 2018.

\bibitem{2}
Rodney Van~Meter.
\newblock {\em Quantum networking}.
\newblock John Wiley \& Sons, 2014.

\bibitem{planning2}
Davide Venturelli, Minh Do, Eleanor~G Rieffel, and Jeremy Frank.
\newblock Temporal planning for compilation of quantum approximate optimization
  circuits.
\newblock In {\em IJCAI}, pages 4440--4446, 2017.

\bibitem{IBM_intro}
Robert Wille, Austin Fowler, and Yehuda Naveh.
\newblock Computer-aided design for quantum computation.
\newblock In {\em 2018 IEEE/ACM International Conference on Computer-Aided
  Design (ICCAD)}, pages 1--6. IEEE, 2018.

\bibitem{Astar}
Alwin Zulehner, Alexandru Paler, and Robert Wille.
\newblock An efficient methodology for mapping quantum circuits to the ibm qx
  architectures.
\newblock {\em IEEE Transactions on Computer-Aided Design of Integrated
  Circuits and Systems}, 2018.

\end{thebibliography}
\end{document}